\documentclass[a4paper,11pt]{article}
\pdfoutput=1 

\usepackage{jheppub} 

\usepackage[T1]{fontenc} 

\usepackage{CJK}
\usepackage{mathtools}
\usepackage{tikz}
\usepackage{amssymb}
\usepackage{epsfig}
\usepackage{bbm}
\usepackage{bbold}
\usepackage{tabularx,booktabs}
\usepackage{graphicx}
\usepackage{hyperref}
\usepackage{romannum}
\usepackage{graphics,graphpap}
\usepackage{graphicx}
\usepackage{amsmath,exscale,amsthm,bm}
\usepackage{graphicx}
\usepackage{yfonts}
\usepackage{mathrsfs}
\usepackage[toc]{appendix}
\usetikzlibrary{trees}
\usetikzlibrary{positioning,arrows}
\usetikzlibrary{decorations.pathmorphing}
\usetikzlibrary{decorations.markings}
\usetikzlibrary{decorations.text}
\usetikzlibrary{fit,chains,calc,shapes.geometric,intersections}
\usepackage{natbib}
\setcitestyle{square,comma,numbers,sort&compress}
\usepackage{circledsteps}
\def\mc#1{\mathcal#1}

\begin{document}
\title{\boldmath\huge Gravitational waves from phase transitions and cosmic strings in neutrino mass 
models with multiple majorons}

\author[]{Pasquale Di Bari,} 
\author[]{Stephen F. King,} 
\author[]{and Moinul Hossain Rahat} 

\affiliation[]{School of Physics and Astronomy, University of Southampton,\\ Southampton, SO17 1BJ, U.K.}

\abstract{
We explore the origin of Majorana masses within the majoron model and how this can lead to the generation of a distinguishable primordial stochastic background of gravitational waves. We first show how in the simplest majoron model only a contribution from cosmic string can be within the reach of planned experiments. We then consider extensions containing multiple complex scalars, demonstrating how in this case a spectrum comprising contributions from both a strong first order phase transition and cosmic strings can naturally emerge. We show that the interplay between multiple scalar fields can amplify the phase transition signal, potentially leading to double peaks over the wideband sloped spectrum from cosmic strings. We also underscore the possibility of observing such a gravitational wave background to provide insights into the reheating temperature of the universe. We conclude highlighting how the model can be naturally combined with scenarios addressing the origin of matter of the universe, where baryogenesis occurs via leptogenesis and a right-handed 
neutrino plays the role of dark matter.
}

\maketitle
\flushbottom

\def\a{\alpha}
\def\b{\beta}
\def\c{\chi}
\def\d{\delta}
\def\e{\epsilon}
\def\f{\phi}
\def\g{\gamma}
\def\h{\eta}
\def\i{\iota}
\def\j{\psi}
\def\k{\kappa}
\def\la{\lambda}
\def\m{\mu}
\def\n{\nu}
\def\o{\omega}
\def\p{\pi}
\def\q{\theta}
\def\r{\rho}
\def\s{\sigma}
\def\t{\tau}
\def\u{\upsilon}
\def\x{\xi}
\def\z{\zeta}
\def\D{\Delta}
\def\F{\Phi}
\def\G{\Gamma}
\def\J{\Psi}
\def\L{\Lambda}
\def\O{\Omega}
\def\P{\Pi}
\def\Q{\Theta}
\def\S{\Sigma}
\def\U{\Upsilon}
\def\X{\Xi}

\def\ve{\varepsilon}
\def\vf{\varphi}
\def\vr{\varrho}
\def\vs{\varsigma}
\def\vq{\vartheta}

\newcommand{\vev}[1]{\langle #1 \rangle}
\def\dg{\dagger}                                     
\def\ddg{\ddagger}                                   
\def\wt#1{\widetilde{#1}}                    
\def\mt{\widetilde{m}_1}
\def\mti{\widetilde{m}_i}
\def\mtj{\widetilde{m}_j}
\def\rt{\widetilde{r}_1}
\def\mtt{\widetilde{m}_2}
\def\mttt{\widetilde{m}_3}
\def\rtt{\widetilde{r}_2}
\def\mb{\overline{m}}
\def\VEV#1{\left\langle #1\right\rangle}        
\def\be{\begin{equation}}
\def\ee{\end{equation}}
\def\ds{\displaystyle}
\def\ra{\rightarrow}

\def\bea{\begin{eqnarray}}
\def\eea{\end{eqnarray}}
\def\NO{\nonumber}
\def\Bar#1{\overline{#1}}
\def\ylz{\textcolor{black}}

\section{Introduction} 

The discovery of gravitational waves (GWs) \cite{LIGOScientific:2016aoc} 
opens new opportunities to test physics beyond the standard model (BSM). This is particularly interesting 
for those models currently
evading constraints from colliders and, more generally, from laboratory experiments.
Even though GWs have so far  been detected only from astrophysical sources, there are many
different processes in the early universe that could lead to the production of detectable primordial stochastic GW backgrounds. 
In particular, a production from the vibration of cosmic strings \cite{Vilenkin:1984ib}
and from strong first order phase transitions \cite{Witten:1984rs,Hogan:1986qda,Turner:1990rc}
provide quite realistic and testable mechanisms  within various extensions of the standard model (SM)
\cite{Fu:2022eun,King:2021gmj,King:2020hyd}.

These two GW production mechanisms are usually studied separately. In this paper we 
show how within the majoron model \cite{Chikashige:1980ui}, an extension of the SM 
explaining neutrino masses and mixing,  a GW spectrum is produced where both sources can give a non-negligible contribution and fall within the sensitivity of planned experiments. In the majoron model a type-I seesaw \cite{Minkowski:1977sc,Yanagida:1979as,GellMann1979, Glashow:1979nm,Mohapatra:1979ia} Lagrangian results as the outcome of a global $U(1)_L$ spontaneous symmetry breaking and Majorana masses are generated by the vacuum expectation value (VEV) of a single complex scalar field. The massless Goldstone boson, identified as the imaginary part of the complex scalar, is dubbed as majoron. The model can also nicely
embed leptogenesis for the explanation of the matter-antimatter asymmetry of the universe~\cite{Fukugita:1986hr}. 

The idea that a strong first order electroweak phase transition 
associated to the lepton number symmetry breaking can generate a stochastic GW background has 
been explored in \cite{Addazi:2019dqt}. In this case a coupling of the complex scalar to the SM Higgs field was considered. A phase transition within the dark sector of the majoron model, disconnected from the electroweak phase transition, was considered in  \cite{Addazi:2020zcj}, where non-renormalisable operators and explicit symmetry breaking terms  have been included in order to enhance the signal. Moreover, a low-scale phase transition, in the keV-MeV range was also considered
in order to reproduce the NANOGrav putative signal at very low frequencies ($\sim 10^{-9}\,{\rm Hz}$) \cite{Arzoumanian:2020vkk}.

A first order phase transition from $U(1)_L$-symmetry breaking in the dark sector, with no coupling of the complex scalar field to the SM Higgs field, was also considered in \cite{DiBari:2021dri} without resorting either to explicit symmetry breaking terms
or to non-renormalizable operators.  Both the case of low and high scale phase transition were explored.  
It was found that at low scales the NANOGrav result
cannot be explained, unless one invokes some enhancement from some unaccounted new effect. On the other hand,
it was found that at high energy scales the  signal can be sufficiently large to fall within the sensitivity of future experiments
such as 
$\mu$Ares~\cite{Sesana:2019vho}, 
DECIGO~\cite{Kawamura:2019jqt}, 
AEDGE~\cite{Bertoldi:2019tck}, 
AION~\cite{Badurina:2019hst}, 
LISA \cite{Caprini:2015zlo}, 
Einstein Telescope (ET) \cite{Hild:2010id},
BBO \cite{Yagi:2011wg}
and CE \cite{LIGOScientific:2016wof}.
However, this result relied on the introduction of an external auxiliary real scalar field
undergoing its own phase transition occurring prior to the complex scalar field phase transition. Once the auxiliary scalar gets a VEV, its mixing with the complex scalar field generates a zero-temperature barrier described by a cubic term in the effective potential of the latter, leading to a strong first order phase transition and detectable GW spectrum. 

In this paper, we show how the role of the auxiliary field can be nicely played by a second complex scalar in a multiple majoron model. We discuss neutrino mass models with spontaneous breaking of multiple global lepton number symmetries, typically with hierarchical scales. The three right-handed (RH) neutrino masses are then generated by different complex scalars each undergoing 
its own independent phase transition occurring, in general, at different energy scales and breaking lepton number along a specific direction
in flavour space. We have, then, what could be referred to as a (RH neutrino) flavoured majoron model. Importantly, we show that a contribution from the
vibration of cosmic strings generated from the spontaneous breaking of the global lepton number symmetry has also to be taken into account to derive the GW spectrum of these models. The overall spectrum then is the sum
of contributions from both production mechanisms: a contribution from strong first order phase transitions and a contribution from the vibration 
of cosmic strings. For sufficiently strong phase transitions, the resultant signal looks like one or more peaks (from phase transition) over a slanted plateau (from cosmic string).  

The paper is organised as follows. In Section 2 we review the traditional single majoron model where the right-right Majorana
mass term, with three RH neutrinos, is generated by a single complex scalar field breaking total lepton number symmetry. 
The differences in the Majorana masses are then to be ascribed to different couplings. 
Even in this traditional setup we point out that a GW production from the vibration of cosmic strings, not accounted for 
in previous works, should be considered and can give a detectable signal. In Section 3 we extend the model with an additional complex scalar with its respective global lepton number symmetry, whose spontaneous breaking gives mass to the two lighter RH neutrinos.
In this way only two distinguished phase transitions occur with hierarchical energy scales.  
We show that the resulting GW spectrum is, in general, the sum of two contributions, one from the lower scale phase transition and one from
the vibration of cosmic strings created at the highest scale symmetry breaking. The corresponding phase transition does not produce a sizeable contribution to the GW spectrum,
but it results into a VEV of the complex scalar field that generates a term entering 
the effective potential describing the second phase transition at a lower scale. This term strongly enhances 
the production of GWs during the second phase transition. In this way the high scale complex scalar associated with the majoron field  provides the external
auxiliary scalar that had to be assumed in \cite{DiBari:2021dri}, so that the model is self-contained and does not rely on
external assumptions. Finally, in Section 4 we consider the case when all three RH neutrino masses are associated with different complex scalars, each charged under a different global lepton number symmetry. 
At high temperatures one has the restoration of a $U(1)_{L_1} \times U(1)_{L_2} \times U(1)_{L_3}$ symmetry. 
While the temperature decreases, a sequential breaking of each $U(1)_{L_I}$ symmetry occurs at a different scale accompanied by a different phase transition. In this case we show that the GW spectrum now 
can receive a contribution from both the two lower scale phase transitions and still from the vibration of cosmic strings at the highest scale symmetry breaking. We show that such a spectrum may have twin peaks from phase transition signals over a slightly sloped plateau of the cosmic string signal. We draw conclusions in Section 5 and point out that the GW spectrum of the model can provide us important information about the reheating temperature of the universe, and that the model fits naturally within a unified framework of solving the puzzles of baryon asymmetry and dark matter.

\section{Primordial GW stochastic background in the single majoron model} 

In this section, we first review the main features of the single majoron model and then discuss
the generation of a stochastic background of primordial GWs.

\subsection{The single majoron model}

The traditional single majoron model is a simple extension of the SM~\cite{Chikashige:1980ui}, where the
spontaneous breaking of a global $U_L(1)$ symmetry generates a Majorana mass term for the RH neutrinos.
The SM field content is then augmented with $N$ RH neutrino fields $N_I \; (I=1,2,\dots,N)$ 
and a complex scalar singlet,
\be
\textcolor{black}{\phi ={1 \over \sqrt{2}}\,\varphi\ e^{i\theta}\,  ,}
\ee
 where the real component is $C\!P$-even and the imaginary component  is $C\!P$-odd. 
 The new scalar $\phi$ has a tree level potential $V_0(\phi)$. 
 For definiteness, we consider the well motivated case $N=3$.
 The tree-level extension of the SM Lagrangian is then given by
 \bea\label{eq:L_l}
-{\cal L}_ {N_I+\phi} & = & 
\textcolor{black}{\left( \overline{L_{\a}}\,h_{\a I}\, N_{I}\, \widetilde{\Phi} 
+  {\lambda_{I}\over 2}  \, \phi \, \overline{N_{I}^c} \, N_{I}
+ {\rm h.c.}\right) + V_0(\phi) \,,}
\eea
where $\widetilde{\Phi}$ is the dual Higgs doublet. In the early universe, above a critical temperature $T_{\rm c}$, 
one has $\langle \phi \rangle = 0$ so that the RH neutrinos are massless. 
Moreover, since the lepton  doublets $L_\alpha$ and the RH neutrinos $N_I$ have $L = 1$, and  $\phi$ has $L=-2$,
lepton number is conserved.  Below $T_{\rm c}$, the $U_L(1)$ symmetry is broken and the scalar $\phi$ acquires a vacuum expectation value  
$\langle \phi \rangle = v_0\,{e^{i\theta_0}}/\sqrt{2}$. In this way the RH neutrinos become massive with Majorana masses $M_I = v_0\, \lambda_I/\sqrt{2}$.
This leads to lepton number violation and  small Majorana masses for the SM neutrinos via type-I seesaw mechanism. 
We assume that $T_{\rm c} \gg T_{\rm ew} \sim 100 \, {\rm GeV}$, so that 
the majoron phase transition occurs prior to the electroweak phase transition and, therefore, 
the Majorana mass term is generated before the Dirac mass term.

Let us consider the simple tree level potential
\begin{eqnarray} \label{eq:V_L}
V_0(\phi) = - \mu^2 |\phi|^2 + \lambda |\phi|^4 \,  ,
\end{eqnarray}
where $\lambda$ is real and positive, in a way that the potential is bounded from below, and  $\mu^2$ is real and positive
to ensure  the existence of degenerate nontrivial stable minima
with  $\langle \phi \rangle   = v_0\,e^{i\theta}/\sqrt{2}$ with $0\leq \theta < 2\pi$ and where $v_0 \equiv \sqrt{\mu^2/\lambda}$.
 After spontaneous symmetry breaking, we can rewrite $\phi$ as\footnote{\textcolor{black}{Notice that $J$ is not the imaginary
 part of $\phi$ but rather related to the phase fluctuation around the vev $\langle \phi \rangle$. Analogously, $S$ is not
 the real part of $\phi$ but rather related to the radial fluctuation. This is easy to see if one rewrites the fluctuation
 $\delta \phi$ about the vev $\langle \phi \rangle $ starting from 
 $\phi = {[(\langle\varphi\rangle +\delta\varphi)/\sqrt{2}}]\,e^{i\,(\theta_0 + \delta \theta)}$. One can then easily identify  $v_0 =\langle\varphi\rangle$, $S = \delta\varphi$ and $J=v_0\,\delta\theta$.   This shows that Eq.~(\ref{sigma}) is equivalent to  $\phi= ({e^{i\theta_0}/ \sqrt{2}})(v_0 +S) e^{i{J\over v_0}}$.}}
\be\label{sigma}
\phi ={e^{i\theta_0} \over \sqrt{2}}\,(v_0 +  S + i \, J) \,  ,
\ee 
where  $S$ is a massive  field with $m_S^2= 2 \lambda v_0^2$  and $J$ is  the majoron, a massless Goldstone field. 
Moreover, RH  neutrino masses $M_I = \lambda_I \, v_0/\sqrt{2}$ are generated by the VEV of $\phi$   and these lead to a light neutrino mass matrix given by the (type-I) seesaw formula
\begin{eqnarray} \label{eq:seesaw}
(m_\nu)_{\alpha\beta} = - {v_{\rm ew}^2 \over 2}{h_{\alpha I} h_{\beta I} \over M_I} \,   ,
\end{eqnarray}
where $v_{\rm ew}=246\,{\rm GeV}$ is the standard Higgs VEV.
Notice that the potential in Eq.~(\ref{eq:V_L}) corresponds to a minimal choice where we are neglecting possible mixing terms between the new complex scalar field $\phi$ and the standard Higgs boson. In this way, the phase transition involves only the dark sector, consisting only of 
$\phi$ and the three RH neutrinos.  Moreover, we are not considering non-renormalisable terms, so that the model is UV-complete. 
 
Since all minima are equivalent, one can always redefine $\theta$ in a way that the symmetry is broken along the direction 
$\theta_0 =0$, 
without loss of generality. The minimum of the potential lies along the real axis and, for all purposes, one 
can consider the potential as a function of $\varphi$, so that one has:
\be
V_0(\varphi) = -{1\over 2}\,\mu^2 \, \varphi^2 + {\lambda\over 4}\,\varphi^4 \,  .
\ee

Let us now discuss the generation of a primordial stochastic background of GWs. 
	 There are two possible sources in the majoron model. The first is an associated strong first order phase transition \cite{DiBari:2021dri} that we discuss in the subsection 2.2. The second is the network of cosmic strings generated by the breaking of the global $U(1)_L$ symmetry that we discuss in the subsection 2.3. The latter has not been discussed before within a majoron model, though it is analogous to the $U(1)_{B-L}$ spontaneous symmetry breaking discussed, for example, in \cite{Buchmuller:2013lra,Dror:2019syi, Fornal:2020esl, Bosch:2023spa,Blasi:2020wpy}.

\subsection{Stochastic GW background from first order phase transition} \label{GWFOPT}

The scalar field and the three RH neutrinos form what we refer to as the dark sector. The dark sector
interacts with the SM sector only via the Yukawa interactions. 
In the early universe finite temperature effects need to be taken into account. 
They will drive  a phase transition, occurring in the dark sector, from the metastable 
vacuum at $\phi = 0$, where lepton number is conserved and RH neutrinos are massless,
to the true stable vacuum at $\phi = v_0/\sqrt{2}$, where lepton number is non-conserved and RH
neutrino are massive (for a recent review on early universe phase transitions and GWs see \cite{Athron:2023xlk}).  
They are described in terms of a finite-temperature effective potential $V_{\rm eff}^T(\phi)$. 
At temperatures above a critical temperature $T_{\rm c}$, finite temperature effects will induce symmetry restoration 
\cite{Kirzhnits:1972ut}.  When temperature drops down the critical temperature, the phase transition occurs
and, in the zero temperature limit, the tree-level potential $V_0(\phi)$ is recovered, in the broken symmetry phase.\footnote{Notice that the reheating temperature of the universe $T_{\rm RH}$ needs
to be higher than $T_{\rm c}$ for both symmetry restoration and symmetry breaking to occur. If it is lower,
the universe history starts directly in the broken phase and there is no phase transition. 
For this reason, finding evidence for a phase transition  and establishing the value of $T_{\rm c}$ would straightforwardly place a lower bound on $T_{\rm RH}$.}

The  finite-temperature effective potential can be calculated perturbatively at one-loop \cite{Dolan:1973qd} 
and is given by the sum of three terms,
\be
V_{\rm eff}^T(\phi) \simeq V_0 (\phi) + V^0_1(\phi) +  V^T_1(\phi) \,  ,
\ee
where the zero-temperature  one-loop contribution $V^0_1(\phi)$ is given by the 
Coleman-Weinberg potential. This can be written, using cut-off regularization, as~\cite{Dolan:1973qd,Anderson:1991zb,Dine:1992wr,Quiros:1999jp}
\bea\label{V01}
V^0_1(\phi) & = & {1 \over 64 \, \pi^2}\,\left\{m_\phi^4 (\phi) \, \left(\log {m^2_\phi(\phi) \over m^2_\phi(v_0)} - {3 \over 2}\right) + 2\,m_\phi^2 (\phi)\,m^2_\phi(v_0) \right. \\ \nonumber
& & \left. \;\;\;\;\;\;\; - 2\,\sum_{I=1,2,3} \, \left[M_I^4(\phi) \, \left(\log {M_I^2(\phi) \over M^2_I(v_0)} - {3 \over 2}\right) + 2\,M^2_I (\phi)\,M_I^2(v_0) \right] \right\}  \,  .
\eea
The pre-factor of two in the second line accounts for two degrees of freedom for each  RH neutrino species. 
The one-loop thermal potential is given by~\cite{Anderson:1991zb,Dine:1992wr,Quiros:1999jp}
\be \label{eq:V_th}
V^T_1(\phi)  = \frac{T^4}{2\pi^2} \left[ J_B\left(\frac{m_\phi^2(\phi)}{T^2}\right) -
2 \sum_I J_F\left(\frac{M_I^2(\phi)}{T^2}\right) \right]  \,  ,
\ee
where the thermal functions are 
\be
J_{B,F}(x^2) = \int_0^{\infty} dy \, y^2 \, \log (1 \mp e^{-\sqrt{x^2+y^2}})\, .
\ee
 The functions $m_\phi^2(\phi)$ and $M_I^2(\phi)$ are the shifted masses given by
 \be
m_\phi^2(\varphi) \equiv {d^2V^0(\varphi)\over d \varphi^2}  = - \lambda v_0^2  + 3 \lambda \varphi^2 
\ee
and
\be
M_I^2(\varphi) = \lambda_I^2 \, {\varphi^2 \over 2} \,  ,
\ee
where we specialized their dependence as a function of $\varphi$ since, even when thermal effects are included, all the study of the dynamics can be done along the real axis of $\phi$ without loss of generality.  

This time the sum over the RH neutrino species, the only fermions coupling to $\phi$, 
should only include those that are fully thermalised prior to the phase transition, while we can neglect the contribution 
from those that are not.  RH neutrinos thermalise at a temperature
\cite{Garbrecht:2013bia,DiBari:2019zcc} 
\be\label{Teq}
T^{\rm eq}_I \simeq 0.2 \, {(h^\dagger \, h)_{II} \, \bar{v}_{\rm ew}^2 \over m_{\rm eq}} \,  ,
\ee  
where $m_{\rm eq} \equiv [16\pi^{5/2}\sqrt{g^\star_\rho}/(3\sqrt{5})]\,(\bar{v}_{\rm ew}^2/M_{\rm P}) \simeq 1.1 \,{\rm meV}\,\sqrt{g^\star_\rho/g^{\rm SM}_\rho}$ 
is the usual effective equilibrium neutrino mass and we simply defined 
$\bar{v}_{\rm ew} = v_{\rm ew}/\sqrt{2} \simeq 174\,{\rm GeV}$. Note that the quantity 
$\overline{M} = \bar{v}_{\rm ew}^2 / m_{\rm eq}
= (3\sqrt{5})/(16\pi^{5/2}\,\sqrt{g^\star_\rho/g^{\rm SM}_\rho})\,M_{\rm P}
\simeq 3\times 10^{16}\,{\rm GeV}\sqrt{g^{\rm SM}_\rho/ g^\star_\rho}$ 
is independent of the electroweak scale $v_{\rm ew}$.
The condition for the thermalisation of the RH neutrino species $N_I$ prior to the phase transition 
can then be written as 
\be\label{hhcond}
(h^\dagger\,h)_{II} \gtrsim 5\,{T_{\rm c} \over \overline{M}} \,  .
\ee 
 The equilibration temperature $T_{\rm eq}$ and the condition Eq.~(\ref{hhcond}) 
 can also be conveniently expressed in terms of the dimensionless RH neutrino decay parameters 
\be
K_{I} = (h^\dagger\,h)_{II}{M_\star \over M_I} \,  ,
\ee
obtaining, respectively,
\be
T_{\rm eq} \simeq 0.2 \, M_I \, K_I \,  \;\;\;\;\; \mbox{\rm and} \;\;\;\;\;
K_I \gtrsim 5 \, {T_{\rm c} \over M_I} \;  .
\ee
Taking into account
the measured values of the solar and atmospheric neutrino mass scales, from the seesaw formula it can be shown that all three RH neutrino species can 
satisfy the condition of thermalisation and this is what we assume for simplicity following \cite{DiBari:2021dri}.\footnote{On the other hand, in the case of a strong hierarchical  RH neutrino spectrum, like in the case of $SO(10)$-inspired models, one can have an opposite situation where
only the heaviest RH neutrino species is fully thermalised prior to the phase transition. One could even have a scenario where no RH neutrino species is thermalised.}
Of course we also assume $T_{\rm RH} \gg T_{\rm c}$ for the phase transition to occur (as noticed in footnote 1).  Another important thermal effect to be taken into account is that the tree-level shifted mass have to be replaced by resummed thermal masses \cite{Parwani:1991gq} 
\be\label{resummation}
m_\phi^2(\varphi) \to {\rm m}_{\phi , T} ^{2}(\varphi) = m_\phi^2(\varphi)    + \Pi_\phi\,  ,
\ee
where the Debye mass $\Pi_\phi$ is given by
\be
\Pi_\phi = \left( \frac{2+d_{\rm scalar}}{12} \lambda + N\,\frac{M^2}{24 v_0^2} \right) T^2 \,  .
\ee 
In this expression one has $d_{\rm scalar} = 2$ for the case of a complex scalar we are considering.  
The quantity $M$ denotes either the mass of the heaviest RH neutrino in the case of hierarchical RH neutrino mass spectrum (in which case $N= 1$), or  a common mass in the case of quasi-degenerate RH neutrinos (in which case $N$ is  the number of RH neutrinos).
This allows us to reduce the number of parameters while spanning the space between $N=1$ (hierarchical RH neutrinos) and $N=3$ (quasi-degenerate RH neutrinos).

With this replacement and neglecting 
${\cal O}((M_I/T)^6)$ terms in the high temperature expansion of the thermal functions,
one obtains the {\it dressed effective potential}~\cite{Curtin:2016urg,Croon:2020cgk,DiBari:2021dri} 
\be\label{VTeffminimal2}
 V^T_{\rm eff}(\varphi) \simeq {1\over 2}\, \widetilde{M}_T^2\,\varphi^2 - A \, T \, \varphi^3 + \frac{1}{4}\lambda_T\, \varphi^4 \,  .
\ee
In this expression we introduced 
\be\label{MTsq}
\widetilde{M}_T^2\equiv 2\,D\,(T^2 - T_0^2) \,   ,
\ee
where $T_0$ is the destabilisation temperature defined by
\be\label{DT0sq}
2\,D\,T_0^2 =   \lambda\,v_0^2 +{N \over 8\,\pi^2}\,{M^4 \over v_0^2} 
-{3\over 8 \,\pi^2}\lambda^2 \, v_0^2  \,   .
\ee
The dimensionless constant coefficients $D$ and $A$ are given by
\be\label{DA}
D = {{\lambda \over 8} + {N\over 24}\,{M^2 \over v_0^2}} \,   \;\;\;\; \mbox{\rm and} \;\;\;\;
A =  {(3\,\lambda)^{3/2} \over 12\pi } \,  \,  .
\ee
Finally, the dimensionless temperature dependent  coefficient $\lambda_T$ is given by
\be\label{lambdaT}
\lambda_T = \lambda  - \frac{N\, M^4}{8\,\pi^2 \, v_0^4}  \, \log {a_F \, T^2 \over e^{3/2}\,M^2} 
+  {9\lambda^2 \over 16 \pi^2}\,\log {a_B \, T^2 \over e^{3/2}\,m_S^2} \,  .
\ee
Notice that one has to impose $M_1 < m_S$ in order for the massive scalar $S$ to decay 
into RH neutrinos in a way that its thermal abundance does not overclose the universe. 
However, this condition is easily satisfied, since the scalar and RH neutrino masses are 
roughly of the same order-of-magnitude as $v_0$.

At very high temperatures the cubic term in the effective potential (\ref{VTeffminimal2}) is negligible
and one has symmetry restoration. However, while temperature drops down, there is a particular 
time when a second minimum at a nonzero value of $\varphi$ forms. When temperature further 
decreases, a barrier separates the two coexisting minima.  The critical temperature $T_{\rm c}$
is defined as that special temperature when the two minima become degenerate. Until this time, the
probability that a bubble of the false vacuum nucleates vanishes but below the critical temperature
it is nonzero. The nucleation probability per unit time and per unit volume can be expressed in terms of the Euclidean action $S_E$ 
as \cite{Coleman:1977py}:
\be
\Gamma(\varphi,T) = \Gamma_0(T)\,e^{-S_E(\varphi,T)} \,  .
\ee
At finite temperatures one has $S_E(\varphi,T) \simeq S_3(\varphi,T)/T$ and 
$\Gamma_0(T) \simeq T^4 \, [S_3(T)/(2\pi\,T)]^{3/2}$~\cite{Linde:1981zj},
where the quantity $S_3$ is the spatial Euclidean action given by
\be\label{S3T}
S_3(\varphi,T) = \int\,d^3x \, \left[ {1 \over 2}\,(\bm{\nabla} \varphi)^2 + V^T_{\rm eff}(\varphi) \right] 
= 4\pi \, \int_0^\infty \,dr \, r^2 \left[ {1 \over 2}\,\left({d^2\varphi \over dr^2}\right)^2 + V^T_{\rm eff}(\varphi) \right] \,  .
\ee
The physical solution for $\varphi$ minimizing $S_3(\varphi,T)$ can be found solving the EoM
\be\label{EoM}
{d^2\varphi \over dr^2} + {2 \over r}\, {d\varphi \over dr} = {d V^T_{\rm eff}(\varphi)   \over dr} \, ,
\ee
with boundary conditions $(d\varphi/dr)_{r=0} = 0$ and $\varphi(r \ra \infty) = 0$. Since
for $T \geq T_{\rm c}$ the nucleation probability vanishes, one has $\lim_{T \ra T_{\rm c}^-} S_E \ra \infty$, while on the 
other hand $\lim_{T \ra T_0} S_E \ra 0$, so that at $T_0$ all space will be in the true vacuum and 
the phase transition comes to its end.\footnote{This is true for not too strong phase transitions, as we will consider, otherwise the Euclidean action might actually reach a minimum and then increase again reaching an
asymptotic non-vanishing value at zero temperature.} 
If the phase transition is quick enough, then one can describe the phase
transition  as occurring within a narrow interval of temperatures about a particular value $T_\star$ 
such that $T_{\rm c} > T_\star > T_0$. The temperature $T_\star$ is referred to as the phase transition temperature and it is usually 
identified with the percolation temperature, defined as the temperature at which the fraction of space still in the  false vacuum is $1/e$. The fraction of space filled by the false vacuum at time $t$  is given by~\cite{Guth:1979bh,Guth:1981uk} 
\be
P(t) = e^{- I(t)} \,  ,
\ee
where 
\be\label{It}
I(t) = {4\pi \over 3} \, \int_{t_{\rm c}}^t \, dt' \,  \Gamma(t') \, a^3(t') \, \left[\int_{t'}^t \, dt'' \, {v_{\rm w} \over a(t'')} \right]^3 \,  ,
\ee
$a(t)$ is the scale factor and $v_{\rm w}$ is the bubble wall velocity. Therefore, $P(t_\star) = 1/e$ corresponds to $I(t_\star) = 1$,
where $t_\star \equiv t(T_\star)$. It can be shown \cite{Megevand:2016lpr} that at $T_\star$ the Euclidean action has to satisfy 
\be\label{SETstar}
S_E(T_\star) - {3\over 2} \, \log {S_E(T_\star) \over 2\pi}   =
  4 \log{T_\star \over H_\star}  - 4\,\log[T_\star \, S'_E(T_\star)]  + 
  \log(8\,\pi \, v^3_{\rm w}) \,  ,
\ee
where $H_{\star} = H(t_{\star})$. This equation allows to calculate $T_\star$ and $S_E(T_\star)$  
having derived $S_E(T)$ from the solution of the EoM. 

The calculation of the GW spectrum produced during the phase transition is characterised by two quantities.
The first is $\beta \equiv \dot{\Gamma}/\Gamma$, the rate of variation of the nucleation rate. Its inverse, $\beta^{-1}$,
gives the time scale of the phase transition. In our case, we are interested in the scenario of fast phase transition, for $\beta^{-1} \ll H^{-1}$,
so that, with a first order expansion of the Euclidean action about $t_{\star}$ 
\be\label{betaoverH}
{\beta \over H_{\star}} \simeq T_{\star} \left.{d(S_3/T) \over dT}\right|_{T_{\star}} \,  .
\ee
This provides a sufficiently good approximation for $\beta/H_{\star} \gtrsim 100$ \cite{Megevand:2016lpr}. 
The second quantity characterising the phase transition is the strength of the phase transition $\a$ defined as 
\be\label{alpha}
\a \equiv {\ve(T_{\star}) \over \rho(T_{\star})} \,  ,
\ee
where $\ve(T_{\star})$ is the latent heat released during the phase transition and $\rho(T_{\star})$ is the
total energy density of the plasma, including both SM and dark sector degrees of freedom.
The latent heat can be calculated using
\be
\ve(T_{\star}) = -\Delta V^{T_\star}_{\rm eff}(\varphi) - T_{\star} \, \Delta s(T_{\star}) =  
-\Delta V^{T_\star}_{\rm eff}(\varphi)  + T_{\star} 
\left.{\partial \Delta V^{T_\star}_{\rm eff}(\varphi) \over \partial T}\right|_{T_{\star}} \,  ,
\ee
where $\Delta V^{T_\star}_{\rm eff}(\varphi) = V^{T_\star}_{\rm eff}(\phi^{\rm true}_1) - V^{T_\star}_{\rm eff}(\phi^{\rm false}_1)$, and in the first relation, from thermodynamics, $\D s$ is the entropy density variation and the free energy of the system has been identified with the effective potential. 
Notice that in our case, $V^{T_\star}_{\rm eff}(\phi^{\rm false}_1)=0$.
Also notice that the constraint  $\beta/H_{\star} \gg 1$ for the validity of 
Eq.~(\ref{betaoverH}) implies a constraint $\alpha \ll 1$, since the two quantities 
are not completely independent of each other with $\beta/H_{\star} \propto \a^{-2}$ \cite{Ellis:2020awk}. 
For definiteness, we will then impose $\alpha \leq 0.3$, corresponding typically to 
$\beta/H_\star \gtrsim 100$.
The total energy density of the plasma can be expressed, as usual, as  
\be
\rho(T) = g_{\rho}(T)\, {\pi^2 \over 30}\,T^4 \,  .
\ee
The number of the total ultrarelativistic degrees of freedom $g_{\rho}(T)$ is in this case given by the sum of 
two contributions, one from the SM and one from the dark sector, explicitly, one has 
$g_{\rho}(T) = g_{\rho}^{\rm SM}(T) + g_{\rho}^{\rm dark}(T)$,
where $g_{\rho}^{\rm SM}(T_\star) =106.75$ and 
$g_{\rho}^{\rm dark}(T_\star) = g_\rho^{\phi} + {7 \over 4} \, N$  with $g_\rho^{\phi} = 2$. 

Let us now calculate the GW spectrum defined as
\be
h^2\O_{{\rm GW}0}(f)=  {1 \over \rho_{{\rm c}0}h^{-2}} \,  {d\rho_{{\rm GW}0}\over d\ln f} \,  ,
\ee 
where $\rho_{{\rm c}0}$ is the critical energy density  and $\rho_{{\rm GW}0}$ is
the  energy density of GW, produced during the phase transition, both calculated at the present time. 
We assume that the phase transition occurs in the detonation regime, i.e., with supersonic bubble wall 
velocities, $v_{\rm w} \geq c_{\rm s} = {1/\sqrt{3}}$, that is typically verified in the regime $\a \leq 0.3$
we are considering. Moreover,
the dominant contribution to the GW spectrum typically comes from sound waves in the plasma, \textcolor{black}{with a sub-dominant contribution from magnetohydrodynamic (MHD) turbulence,
so that $h^2 \, \O_{{\rm GW} 0}(f) = h^2 \, \O_{\rm sw 0}(f) + h^2 \, \O_{\rm tb 0}(f) \simeq h^2 \, \O_{\rm sw 0}(f)$ \cite{Caprini:2015zlo}}. 

A numerical fit to the the GW spectrum that is the result of semi-analytical methods and at the same time
takes into account the results of numerical simulations, quite reliable in the regime $\alpha \leq 0.3$ we are considering, yields~\cite{Caprini:2015zlo,Hindmarsh:2017gnf,Cutting:2019zws}
\be\label{omegasw}
h^2\Omega_{\rm sw 0}(f) =3 \, h^2 \, r_{\rm gw}(t_\star,t_0)\,\widetilde{\Omega}_{\rm gw} \,  
H_\star \, R_\star \, \left[\frac{\kappa(\a)\, \alpha}{1+\alpha}\right]^2 \, 
\widetilde{S}_{\rm sw} (f) \, \Upsilon(\alpha,\beta/H_{\star})\, ,
\ee
where the redshift factor $r_{\rm gw}(t_\star,t_0) $, evolving $\Omega_{\rm gw\star} \equiv \rho_{\rm gw\star}/\rho_{{\rm c}\star}$ into $\Omega_{\rm gw 0} \equiv \rho_{\rm gw 0}/\rho_{\rm c 0}$, is given by 
\cite{Kamionkowski:1993fg}
\be
r_{\rm gw}(t_\star,t_0) = \left({a_\star \over a_0}\right)^4 \, \left({H_\star \over H_0}\right)^2 
= \left({g_{S0}\over g_{S\star}}\right)^{4 \over 3}\,{g_{\r\star} \over g_\gamma} \, \O_{\gamma 0}
 \simeq 3.5 \times 10^{-5} \, \left({106.75 \over g_{\rho \star}} \right)^{1\over 3} \,  \left({0.6875 \over h}\right)^2 ,
\ee
and in the numerical expression we used: $g_\gamma =2$, $g_{S\star} = g_{\rho\star} $,
$g_{S0} = 43/11 \simeq 3.91$ , $\O_{\gamma 0} = 0.537\times 10^{-4} (0.6875/h)^2$.
Replacing the expression for the mean bubble separation $R_\star = (8\pi)^{1/3} v_{\rm w}/\beta$, valid in the detonation regime we are assuming, we obtain the numerical expression 
\be\label{omegasw2}
h^2\Omega_{\rm sw 0}(f) = 1.45 \times 10^{-6} \, \left({106.75 \over g_{\rho \star}} \right)^{1\over 3} \,
\left(\widetilde{\Omega}_{\rm gw} \over 10^{-2} \right)\, \left[\frac{\kappa(\a)\, \alpha}{1+\alpha}\right]^2 \, 
{v_{\rm w} \over \beta/H_{\star}}\,\widetilde{S}_{\rm sw} (f) \, \Upsilon(\alpha,\beta/H_{\star}) \,  .
\ee
The normalised spectral shape function is given by $\widetilde{S}_{\rm sw} (f) \simeq 0.687\, S_{\rm sw} (f)$ with
\begin{eqnarray}\label{Ssw}
S_{\rm sw} (f) = \left(\frac{f}{f_{\rm sw}}\right)^3 \left[\frac{7}{4+3({f/f_{\rm sw}})^2} \right]^{7/2} \,  ,
\end{eqnarray} 
where $f_{\rm sw}$ is the peak frequency given by
\begin{eqnarray} \label{fpeak}
f_{\rm sw} =8.9\,\mu{\rm Hz} \, \frac{1}{v_{\rm w}} \frac{\beta}{H_\star} \left( \frac{T_\star}{\rm 100\,GeV}\right) \left( \frac{g_{\rho\star}}{106.75} \right)^{1/6} \, .
\end{eqnarray}
Notice that we have normalized the number of degrees of freedom to the SM value since we are discussing phase transitions at or above the electroweak scale.
The efficiency factor $\kappa(\a)$ measures how much of the vacuum energy is converted to bulk kinetic energy. 
We adopt Jouguet detonation solutions since we assume that the plasma velocity behind the bubble wall
is equal to the speed of sound. Then, the efficiency factor is ~\cite{Steinhardt:1981ct,Espinosa:2010hh} 
\begin{eqnarray}
\kappa(\a) \simeq {\alpha\over 0.73+0.083\sqrt{\alpha}+\a} \,,
\end{eqnarray}
and the bubble wall velocity is $v_{\rm w}(\a) = v_{\rm J}(\a)$, where
\begin{eqnarray} \label{eq:Jouguet}
v_{\rm J}(\a) \equiv \frac{\sqrt{1/3} + \sqrt{\alpha^2 +2\alpha/3}}{1+\alpha}\,.
\end{eqnarray}
Jouguet solutions provide a simple  prescription but a rigorous description would require
numerical solutions of the Boltzmann equations~\cite{Espinosa:2010hh}. 
The prefactor $\widetilde{\Omega}_{\rm gw}$ in Eq.~(\ref{omegasw}) is calculated from numerical simulations and 
a recent analysis shows that in the regime we are considering, for $\alpha \leq 0.3$
and $v_{\rm w} = v_{\rm J} \gtrsim c_{\rm s}$, it takes values approximately
in the range $\widetilde{\Omega}_{\rm gw}= 10^{-3}$--$10^{-2}$ \cite{Cutting:2019zws},
with the exact value depending on  additional 
parameters necessary to simulate the GW production from sound waves, such as friction,
that we do not describe in our analysis.  For this reason we show in all results
bands of GW spectra corresponding to this range of values for $\widetilde{\Omega}_{\rm gw}$ rather than a single curve.  This should also account for the  use of simple Jouguet solutions
for $v_{\rm w}$ rather than solutions of Boltzmann equations, also depending on friction as
additional parameter. 

Finally, notice that in Eq.~(\ref{omegasw}) there is also a suppression factor $\Upsilon(\alpha,\beta/H_\star) < 1$  which decreases with the strength of the phase transition and is given by 
\cite{Ellis:2018mja,Guo:2020grp}:
\be
\Upsilon(\alpha,\beta/H_\star) =  1- {1\over \sqrt{1+ 2\, H_{\star} \tau_{\rm sw}}} \,  ,
\ee
where the product of the lifetime of the sound waves $\tau_{\rm sw}$ with the Hubble expansion parameter
at the time of the phase transition can, in turn, be
expressed in terms of $\alpha$ and $\beta/H_{\star}$ as
\be
H_\star \tau_{\rm sw}  =  (8\,\pi)^{1\over 3}{v_{\rm w}\over \beta/H_{\star}} \left[ {1 + \a \over \kappa(\a) \, \alpha}\right]^{1/2} \,  .
\ee

\textcolor{black}{For the MHD turbulence contribution, the GW spectrum is given by \cite{Caprini:2015zlo}
\begin{align}
	h^2\ \Omega_{\rm tb 0}(f) &= 3.28 \times 10^{-4}  \left(\frac{106.75}{g_\star}\right)^{1/3} 
\left( \frac{\kappa_{\rm tb}(\alpha)\ \alpha}{1+\alpha} \right)^{3/2} \frac{v_{\rm w}}{\beta/H_\star}\ S_{\rm tb}(f),
\end{align}
where $\kappa_{\rm tb}(\alpha) = \epsilon \kappa(\alpha)$, with $\epsilon \approx 0.05$ representing the fraction of bulk motion of the plasma which is turbulent. The other quantities are given by
\begin{align}
	S_{\rm tb} &= \left(\frac{f}{f_{\rm tb}} \right)^3\ \left[1+ \frac{f}{f_{\rm tb}}\right]^{-11/3}\ \left(1+8\pi f/h_\star\right)^{-1}, \\
	h_\star &= 16.68\ \mu {\rm Hz} \left(\frac{T_\star}{100\ {\rm GeV}}\right)\ \left(\frac{g_\star}{106.75}\right)^{1/6}, \\
	f_{\rm tb} &= 27.3\ \mu {\rm Hz} \frac{1}{v_{\rm w}} \left( \frac{\beta}{H_{\star}} \right) \left(\frac{T_\star}{100\ {\rm GeV}}\right)\ \left(\frac{g_\star}{106.75}\right)^{1/6}.
\end{align}
}

Let us now calculate the GW spectrum within the majoron model.
If we consider the minimal tree level potential in Eq.~(\ref{eq:V_L}), there is a simple solution of the EoM for the 
Euclidean action given by \cite{Dine:1992wr,DiBari:2021dri}
\be\label{euclidean}
{S_3 \over T} = {\widetilde{M}_T^3 \over A^2 \, T^3} \, f(a) \,  ,
\ee
where we defined the dimensionless parameter
\be
a \equiv {\lambda_T \, \widetilde{M}_T^2 \over 2\,A^2 \, T^2} \,  .
\ee
and where
\be
f(a)  \simeq 4.85\,\left[1 +{a \over 4}\,\left(1 +{2.4 \over 1-a} + {0.26 \over (1-a)^2} \right)\right] \,   
\ee
provides an accurate analytical fit.  Using this expression for the Euclidean action, for a given choice of the model
parameters $v_0,\lambda$ and $M$, one can calculate the critical temperature using
Eq.~(\ref{SETstar}). From this one can calculate the parameters $\alpha$ and $\beta/H_\star$ 
and then finally derive the GW spectrum from Eq.~(\ref{omegasw}).

\begin{table}
\centering
{\renewcommand{\arraystretch}{1.2}
\begin{tabular}{c@{\hskip 0.1in}c@{\hskip 0.1in}c@{\hskip 0.1in}c@{\hskip 0.1in}|c@{\hskip 0.1in} c@{\hskip 0.1in}c@{\hskip 0.1in}c@{\hskip 0.1in}c}
    \toprule

  B.P.  & $\lambda$ & $v_0$ [GeV] & $M$ [GeV] & $\alpha$ & $\beta/H_\star$ & $T_{\star}$ [GeV] & $\langle \phi\rangle^{\rm true}_{T_{\star}}$ [GeV] \\
 \midrule
 solid & $2$ & $10^{15}$  & $10^{15}$ & $0.00019$ & $219.9$ & $1.68 \times 10^{15}$ & $5.7 \times 10^{14}$ \\
 dashed & $2$ & $10^{14}$  & $10^{14}$ & $0.0017$ & $1002.1$ & $1.70 \times 10^{14}$ & $5.39 \times 10^{13}$ \\
 dotted & $2$ & $10^{3}$  & $10^{3}$ & $0.0015$ & $65895.5$ & $1713.5$ & $502.6$\\
\bottomrule
\end{tabular}
}
\caption{Benchmark points for gravitational wave signals from first order phase transition of $\phi$. }
\label{table:BP0}
\end{table}

\begin{figure}
\begin{center}
\psfig{file=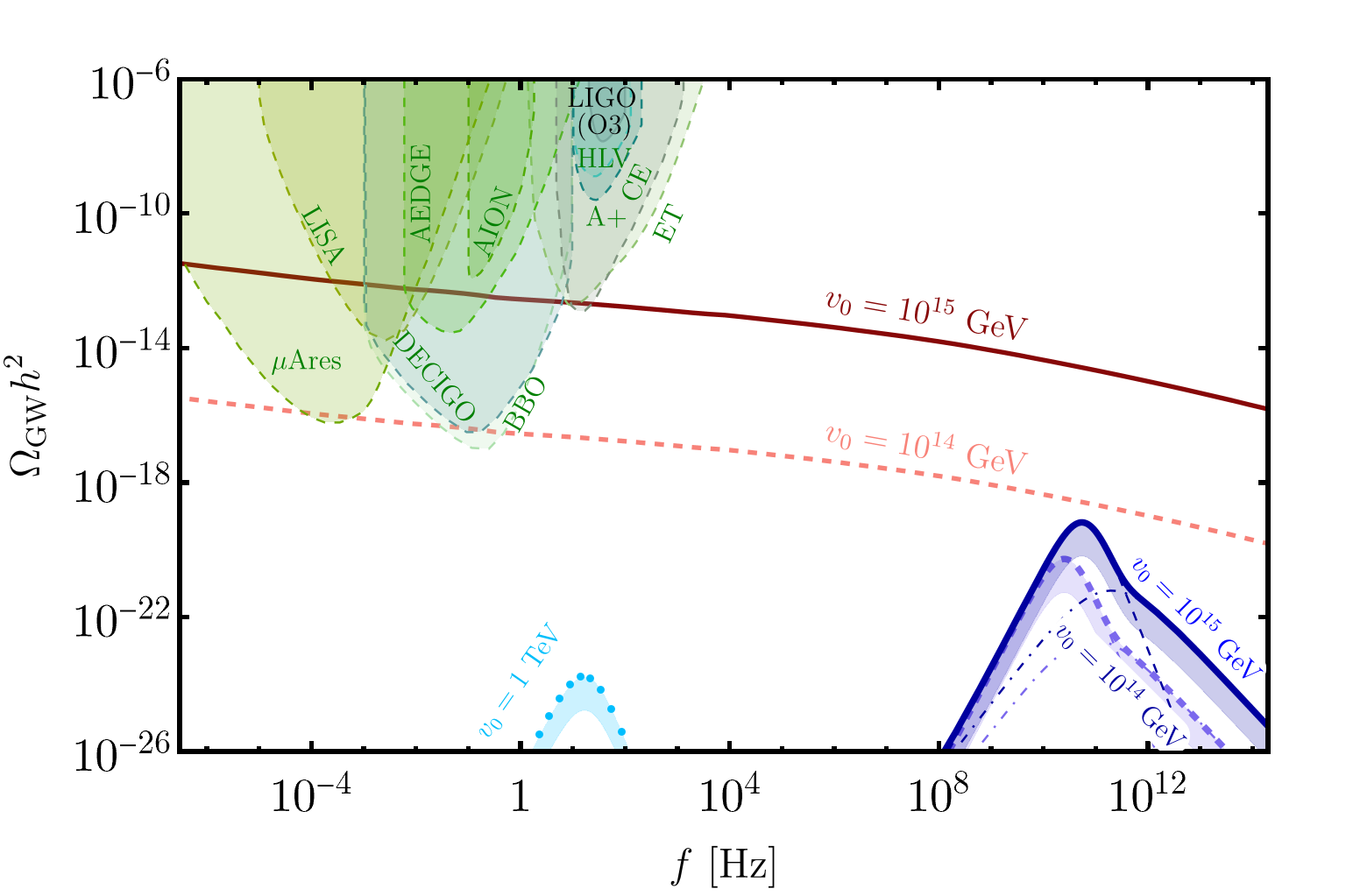,width=0.99\textwidth} 
\end{center}
\caption{The blue bands denote the contribution to the predicted GW spectrum from first order phase transition for $v_0 = 10^{15}\,{\rm GeV}$ (solid line), $10^{14}\,{\rm GeV}$ (dashed line) and $1\,{\rm TeV}$ (dotted line). The red lines denote the contribution from cosmic strings for $v_0 = 10^{15}\,{\rm GeV}$ (solid line) and $10^{14}\,{\rm GeV}$ (dashed line), the signal for $v_0=1$ TeV is too suppressed to show here. The (green) shadowed regions show
the sensitivity curves of the indicated experiments.}
\label{f1}
\end{figure}

In Fig.~\ref{f1} we show, with blue bands, the GW spectra corresponding to the three benchmark choices for the  values of    $v_0,\lambda$ and $M$ in Table~1. \textcolor{black}{The thin dashed and dot-dashed lines correspond to sound wave and MHD turbulence contributions to the GW spectra, whereas the thick lines represent the combined spectra. The contribution from sound wave is dominant at the peak amplitude, while the MHD turbulence modifies the high-frequency tail.}
We also show the sensitivity regions of LIGO \cite{KAGRA:2013rdx, KAGRA:2021kbb} and some planned/proposed experiments, $\mu$Ares \cite{Sesana:2019vho},
LISA \cite{Caprini:2015zlo}, BBO \cite{Yagi:2011wg}, DECIGO \cite{Kawamura:2019jqt}, AEDGE \cite{Bertoldi:2019tck}, AION \cite{Badurina:2019hst},
ET \cite{Hild:2010id} and CE \cite{LIGOScientific:2016wof}.  

Considering that these three choices are those found in a scan that maximise the signal in respective peak frequencies, 
it should be clear that the contribution from phase transitions in the case of the 
minimal model is far below the experimental sensitivity. In this way we confirm the conclusions
found in \cite{DiBari:2021dri}. In the next subsection we point out, however, that at least for large values of $v_0$,
the contribution from cosmic strings could be detectable in future experiments even in this minimal model.

Before concluding we should also mention that one could think to add an explicit symmetry breaking 
cubic term in the tree level potential. However, as noticed in \cite{DiBari:2021dri}, its coefficient is
upper bounded by the observation that it unavoidably generates also a linear term in the effective 
potential. This tends to remove the barrier between the two vacua so that, if the coefficient is too large, 
there is no first order phase transition and, therefore, no GW production. For this reason, we do not 
pursue this scenario.

\subsection{GW from global cosmic strings}

Spontaneously breaking the $U(1)_L$ symmetry at high energies by the complex scalar $\phi$ generates a global cosmic string network, which dominantly radiates Goldstone bosons, and sub-dominantly emits gravitational waves \cite{Vilenkin:1986ku}. Compared to the Nambu-Goto string-induced almost flat gravitational wave spectrum associated with a gauged symmetry breaking, the global cosmic string-induced gravitational waves are typically suppressed, and their amplitude mildly falls off with frequency for most of the spectrum of interest. This makes their detection at interferometers more challenging, unless the symmetry breaking scale $v_0$ is above $10^{14}$ GeV. In this section we briefly review the dynamics of global cosmic strings using  the velocity-dependent one scale (VOS) model \cite{Martins:1996jp, Martins:2000cs, Martins:2003vd, Martins:2016wqq, Martins:2018dqg, Correia:2019bdl} and the associated gravitational wave spectrum  following \cite{Chang:2021afa}.\footnote{We assume that symmetry breaking and the consequent formation 
of the cosmic string network occur after inflation. In the case that the symmetry breaking happens during inflation, the GW spectrum gets modified 
by inflation, see for example \cite{Lazarides:2021uxv} and references therein. We also assume that the cosmic strings are stable. This is always
the case if a matter parity remains unbroken after $U(1)_L$ symmetry breaking (for example, see  discussion in \cite{Dror:2019syi}).}


The global cosmic string network consists of horizon sized long strings that randomly intersect and form sub-horizon sized loops at the intersections. The network shrinks and loses energy with time, but eventually enters a scaling regime where the average inter-string separation scale $L$, and the ratio of the energy density of the network to the total background energy density remain constant. For Nambu-Goto strings, this energy is radiation from the string loops predominantly in the form of gravitational waves. However, for global strings the leading mode of energy radiation is from the emission of Goldstone particles, and only a fraction of the energy is radiated as gravitational waves. 

The energy density of the global string network can be expressed as
\begin{align}
\rho_{\rm cs} = \frac{\mu(t)}{L^2(t)} = \frac{ \mu(t)}{t^2} \xi(t), \label{rho}
\end{align}
where $\mu(t)$ is the energy per unit length of the long strings, and $\xi(t)$ is a dimensionless parameter which represents the number of long strings per horizon volume. While for Nambu-Goto strings, $\mu$ is a constant, for global strings it has a logarithmic dependence on the ratio of two scales, a macroscopic scale $L(t)$ close to the Hubble scale, and a microscopic scale $\delta(t) \sim 1/(\lambda v_0)$ representing the width of the string core, 
\begin{align}
 \mu(t) = 2\pi v_0^2 \log{\frac{L(t)}{\delta}} \equiv 2\pi v_0^2 N(t). \label{mu}
\end{align}
Here we have defined a dimensionless time parameter $N(t) \equiv \log[{L(t)/\delta(t)}]$.
Eq.~\eqref{mu} can then be written as
\begin{align}
N(t) + \frac{1}{2}\log{\xi(t)} = \log{v_0 t}, \label{N}
\end{align}
assuming the quartic coupling $\lambda \sim 1$.

The evolution of the inter-string separation scale $L(t)$ and the average long string velocity $\bar{v}$ are given by a system of coupled differential equations, 
\begin{align}
\left(2-\frac{1}{N}\right) \frac{dL}{dt} &= 2HL\left(1+\bar{v}^2\right) + \frac{L \bar{v}^2}{\ell_f} + \bar{c} \bar{v} + s \frac{\bar{v}^6}{N},  \label{eomL} \\
\frac{d\bar{v}}{dt} &= (1-\bar{v}^2) \left( \frac{\bar{q}}{L} - 2H\bar{v} \right). \label{eomv}
\end{align}
The first term on the RHS of Eq.~\eqref{eomL} represents dilution effect from Hubble expansion. The second term gives a negligible thermal friction effect with a characteristic scale $\ell_f \propto \mu/T^3$. The third term stands for the loop chopping effect, where $\bar{c}$ is the rate of loop chopping. The fourth term represents the backreaction due to the Goldstone emission. 
The quantity $\bar{q}$ is a momentum parameter. In analogy with Nambu-Goto strings, the solution of Eqs.~\eqref{eomL} and \eqref{eomv} can be expressed as
\begin{align}
L^2(t)&= \frac{t^2}{8} \frac{n \bar{q} (\bar{q} + \bar{c}) (1+\Delta)}{1-\frac{2}{n}-\frac{1}{2N(t)}}, \label{Lsol}\\
\bar{v}^2(t) &= \frac{1-\Delta}{2}\frac{n\bar{q}}{\bar{q}+\bar{c}}\left(1-\frac{2}{n}-\frac{1}{2N(t)}\right), \label{vsol}
\end{align}
where $\Delta \equiv \bar{\kappa}/(N(\bar{q}+\bar{c}))$, $\bar{\kappa}\equiv s \bar{v}^5/(1-\Delta)^{5/2}$, and $n= 3, 4$ corresponds to matter and radiation domination, respectively. Fitting data extracted from the simulation results in Refs. \cite{Gorghetto:2018myk, Hindmarsh:2019csc, Klaer:2017qhr}, the VOS model parameters can be approximated as \cite{Chang:2021afa}
\begin{align}
\{\bar{c}, \bar{q}, \bar{\kappa}\} \simeq \{0.497, 0.284, 5.827\}.
\end{align}
Since $\xi(t) = t^2/L^2(t)$ from Eq.~\eqref{rho}, Eq.~\eqref{Lsol} can be used to express $\xi$ as a function of $N(t)$. Eq.~\eqref{N} then expresses $N(t)$ as a function of $t$. Similarly, $\bar{v}$ can be expressed as function of $t$ from Eq.~\eqref{vsol}.

Assuming that the loop size during formation of the string network is given by $\ell_i \sim \alpha t_i$, where 
$\a$ is a dimensionless ${\cal O}(1)$ parameter not to be confused with the strength of the phase transition, and the fraction of energy density of the strings contributing to gravitational wave $F_\alpha \sim 0.1$ \textcolor{black}{(for $\alpha \sim 0.1$)} \cite{Blanco-Pillado:2013qja, Blanco-Pillado:2017rnf}, the formation rate of string loops is given by
\begin{align}
\frac{d\rho_0}{dt} \times \mc{F}_\alpha = - \frac{d\rho_{\rm cs}}{dt} \times F_\alpha \times \mc{F}_\alpha = \mathcal{E}_{\rm loop} \frac{\mu}{t^3} F_\alpha \mc{F}_\alpha, \label{drhodt}
\end{align}
where $\mathcal{F}_\alpha \sim 1$ is the loop size distribution function, and $\mathcal{E}_{\rm loop} \equiv \bar{c} \, \bar{v} \, \xi^{3/2}$ is the loop emission parameter.

After formation, the string loop rapidly oscillates and radiates energy in the form of Goldstone particles and gravitational waves until disappearing completely with a rate \cite{Vilenkin:1986ku}
\begin{align}
\frac{dE}{dt} = -\Gamma G\mu^2 - \Gamma_a v_0^2 \,  , \label{Et} 
\end{align}
where we assume the benchmark values $\Gamma \sim 50$ \cite{Blanco-Pillado:2013qja, Blanco-Pillado:2017oxo, Vilenkin:1981bx, Blanco-Pillado:2011egf} and $\Gamma_a \sim 65$ \cite{Vilenkin:2000jqa, Battye:1997jk}. 
The size of a loop initial length $\ell_i = \alpha \, t_i$ at a later time can be expressed as
\begin{align}
\ell (t) \simeq \alpha\, t_i - \Gamma G \mu (t-t_i) - \frac{\Gamma_a}{2\pi} \frac{t-t_i}{\log{N}}, \label{ell}
\end{align}
where the second and third terms represent the decrease in loop size for gravitational wave emission and Goldstone emission, respectively. 

It is useful to decompose the radiation into a set of normal modes $\tilde{f}_k = 2k/\tilde{\ell}$, where $k=1,2,3, \ldots$, and $\tilde{\ell} \equiv \ell (\tilde{t})$ is the instantaneous size of a loop when it radiates at $\tilde{t}$. Accordingly, the radiation parameters can be decomposed as $\Gamma = \sum_k \Gamma^{(k)}$ and $\Gamma_a = \sum_k \Gamma_a^{(k)}$, where
\begin{align}
\Gamma^{(k)} = \frac{\Gamma k^{-4/3}}{\sum_{j=1}^{\infty} j^{-4/3}}, \quad \text{and\quad} \Gamma_a^{(k)} = \frac{\Gamma_a k^{-4/3}}{\sum_{j=1}^{\infty}j^{-4/3}},
\end{align}
and the normalization factor is approximately $\sum_{j=1}^{\infty}j^{-4/3} \simeq 3.60$. 

Taking redshift into account, the observed frequency at today's interferometers is
\begin{align}
f_k = \frac{a(\tilde{t})}{a(t_0)} \tilde{f}_k,
\end{align}
where $t_0$ is present time and the scale factor today is $a(t_0) \equiv 1$. The relic gravitational wave amplitude is summed over all normal modes
\begin{align}
\Omega_{\rm GW}(f) = \sum_k \Omega_{\rm GW}^{(k)} (f) = \sum \frac{1}{\rho_c}\frac{d\rho_{\rm GW}}{d \log{f_k}}.
\end{align}
From Eqs.~\eqref{drhodt} and \eqref{ell}, the contribution from an individual $k$ mode can be expressed as
\begin{align}
\Omega_{\rm GW}^{(k)} (f) = \frac{\mc F_a F_a}{\alpha \rho_c} \frac{2k}{f} \int_{t_f}^{t_0} d\tilde{t} \frac{\mathcal{E}_{\rm loop}\left(t_i^{(k)}\right)}{t_i^{(k)4}} \frac{\Gamma^{(k)}G\mu^2}{\alpha + \Gamma G \mu + \frac{\Gamma_a}{2\pi N}} \left[\frac{a(\tilde{t})}{a(t_0)}\right]^5 \left[\frac{a\left(t_i^{(k)}\right)}{a(\tilde{t})}\right]^3 \theta(\tilde{\ell}) \theta(\tilde{t}-t_i), \label{GWk}
\end{align}
where $t_f$ is the formation time of the string network. Heaviside theta functions ensure causality and energy conservation. $t_i^{(k)}$ represents the time when a loop is formed, which emits gravitational wave at the time $\tilde{t}$, and is given by
\begin{align}
t_i^{(k)} = \frac{\tilde{\ell}(\tilde{t},f,k) + \left(\Gamma G \mu + \frac{\Gamma_a}{2\pi N}\right)\tilde{t}}{\alpha + \Gamma G \mu + \frac{\Gamma_a}{2\pi N}},
\end{align} 
where the loop size can be written as $\tilde{\ell} = 2\,k\,a(\tilde{t})/f$.

The frequency spectrum of the gravitational wave amplitude is calculated by numerically evaluating Eq.~\eqref{GWk}, and summing from $k= 1$ to a large value to ensure convergence. Evidently, the spectrum can be divided into three regions. The first region corresponds to very high frequencies starting from a cutoff value $f_{v_0}$, where the signal falls off, and the exact shape depends on the initial conditions and very early stages of the string network evolution not fully captured by the VOS model. The cutoff value $f_{v_0}$ is related to the time when the Goldstone radiation becomes significant. In the intermediate radiation dominated region $f_{\rm eq}<f<f_{v_0}$, the spectrum gradually declines as $~\log^3{(\tilde{\ell}^{-1}/f)}$. In the matter dominated region $f_0<f<f_{\rm eq}$, the spectrum behaves as $f^{-1/3}$. The frequency $f_{\rm eq}$ is related to the time of matter-radiation equality, while $f_0$ to the emission at the present time. These characteristic frequencies are given by 
\begin{align}
f_{v_0} &\sim \frac{2}{\alpha t_n} \frac{a(t_{v_0})}{a(t_0)} \sim 10^{10}\ \text{Hz}, \\
f_0 &\sim \frac{2}{\alpha t_0} \sim 3.6 \times 10^{-16}\ \text{Hz}, \\
f_{\rm eq} &\sim 1.8 \times 10^{-7}\ \text{Hz}. 
\end{align}
Although the gravitational wave spectrum from global cosmic strings span over a very wide frequency range, for our purposes we will be concerned mostly in the $\mu$-Hz to kilo-Hz range, where some of the planned interferometers are sensitive. This range falls under $f_{\rm eq} < f < f_{v_0}$. The gravitational wave spectrum can be approximately expressed in this regime by \cite{Chang:2021afa}
\begin{align}
	\Omega_{\mathrm{GW}}(f) h^2 \simeq 8.8 \times 10^{-18}\left(\dfrac{v_0}{10^{15} \mathrm{GeV}}\right)^4 \log ^3\left[\left(\dfrac{2}{\alpha f}\right)^2 \dfrac{v_0}{t_{\mathrm{eq}}} \dfrac{1}{z_{\mathrm{eq}}^2 \sqrt{\xi}} \Delta_R^{1 / 2}(f)\right] \Delta_R(f),
\end{align}   
where $z_{\rm eq} \simeq 8000$ \cite{Aghanim:2018eyx}, and $\Delta_R(f)$ represents the effect of varying number of relativistic degrees of freedom over time:
\begin{align}
\Delta_R(f)=\frac{g_*(f)}{g_*^0}\left(\frac{g_{* S}^0}{g_{* S}(f)}\right)^{4 / 3}.
\end{align} 

\textcolor{black}{We note that we have focused on the GW spectrum from decaying cosmic string loops. Another approach, the Abelian-Higgs model, that takes into account the contribution from long string network is expected to be subdominant \cite{Buchmuller:2013lra}.}

\textcolor{black}{For our numerical calculations we have set $\alpha = 0.1$ as the peak value of the loop sizes at the time of their formation inspired by results from Nambu-Goto string simulations \cite{Blanco-Pillado:2013qja, Blanco-Pillado:2017rnf}. The resulting GW spectrum is modified by up to an order of magnitude if we deviate from this choice \cite{Chang:2021afa}. In the standard radiation-dominated cosmology, $\alpha \lesssim 0.1$ leads to smaller lifetime of the loops, higher string tension, and larger available string energy density to produce GWs. Furthermore, smaller $\alpha$ implies that loops emit GW at higher frequency, hence the amplitude for a given frequency observed today is higher in the frequency range we are interested in. For $\alpha \gg 0.1$, the loops are long-lived and resemble the scenario of Nambu-Goto strings, where the GW amplitude rises with $\alpha$ as $\Omega_{\rm GW} \propto \alpha^{1/2}$. On the other hand, a recent simulation of global cosmic string \cite{Gorghetto:2018myk} suggests a log-normal distribution of $\alpha$, in which case the GW amplitude is also enhanced by a factor of few \cite{Chang:2021afa}.  Our choice of $\alpha \sim 0.1$ can therefore be treated as a conservative choice as far as the GW amplitude is concerned.}

There are several constraints on the global cosmic string formation scale~$\sim v_0$. The dominant radiation mode from global strings is emission of Goldstone bosons. Assuming they remain massless, the upper limit on the total relic radiation energy density from CMB $\Delta N_{\rm eff} \lesssim 0.2$ \cite{Aghanim:2018eyx} implies $v_0 \lesssim 3.5 \times 10^{15}$ GeV \cite{Chang:2021afa}. If we assume standard cosmology, non-observation of gravitational waves at Parkes Pulsar Timing Array (PPTA) \cite{Blanco-Pillado:2017rnf, Lasky:2015lej, Shannon:2015ect} gives an upper bound $v_0 < 2\times 10^{15}$ GeV. Other constraints from inflation scale and CMB anisotropy bound require $v_0 \lesssim \mathcal{O}(10^{15})$ GeV \cite{Chang:2019mza, Lopez-Eiguren:2017dmc}. Hence we consider the global lepton number symmetry violation at scales $\lesssim 10^{15}$ GeV. Furthermore, we require $T_{RH} \gtrsim v_0$ to ensure that the lepton number symmetry is restored in the early universe and symmetry breaking can take place at the scale $\sim v_0$.


We show the global cosmic string induced GW signals for $v_0 = 10^{14}$ and $10^{15}$ GeV in Fig.~\ref{f1} with red curves. The former is within the sensitivity of upcoming interferometers $\mu$Ares, DECIGO and BBO, whereas the latter might be probed at LISA, AEDGE and Einstein Telescope as well. The phase transition signals for  $v_0 = 10^{14}$ and $10^{15}$ GeV remain buried under their respective cosmic string signals.\footnote{If the reheating temperature is below $v_0$, the universe would start in a broken phase, and there would be no signals from either cosmic strings or phase transition.} We therefore conclude that the single majoron model can still be probed in GW interferometers through its cosmic string signal as long as the global lepton number symmetry is spontaneously broken in between $10^{14}$ and $10^{15}$ GeV. 


\section{GW from Majorana mass genesis in a two-majoron model}

As we discussed, the GW contribution to the stochastic background from a phase transition in the single majoron 
model is by far below the sensitivity of planned experiments. 
The reason for the suppressed signal amplitude can be traced back to the fact that for a single scalar, the cubic term is strictly temperature dependent and vanishes at zero temperature. 
It was noticed in \cite{DiBari:2021dri} that
the signal can be strongly enhanced if an auxiliary scalar field is introduced. 
This would undergo its own phase transition  getting its final VEV prior to the phase transition of the original scalar. In this way a bi-quadratic mixing term 
could be added to the tree level potential. This term
generates a zero temperature barrier in the thermal effective potential able to enhance the strength of the phase transition $\alpha$ and, consequently, the GW spectrum.\footnote{This effect has been intensively
employed in electroweak baryogenesis, where the phase transition of the Higgs boson is typically either not taking place at all or too weak, and can be enhanced in the presence of a real auxiliary scalar, which introduces a temperature-independent cubic term to the thermal effective potential of the Higgs field.}
The nature of the auxiliary scalar field was not 
specified in \cite{DiBari:2021dri}. Here we propose a model with two majorons where the auxiliary scalar field
is identified as a complex scalar field charged under a new global lepton number symmetry.

For definiteness, we call the complex scalars $\phi_1$, $\phi_3$, and their respective global lepton number symmetries  $U(1)_{L_{1}}$, $U(1)_{L_3}$. The Lagrangian can be written as ($I=1,2,3$)
\begin{align}\label{eq:L_2}
-{\cal L}_ {N_I+\phi_1+\phi_3} & =  
 \left(\overline{L_{\a}}\,h_{\a I}\, N_{I}\, \widetilde{\Phi} 
+  {y_{1}\over 2}  \, \phi_1 \, \overline{N_{1}^c} \, N_{1} +  {y_{2}\over 2}  \, \phi_1 \, \overline{N_{2}^c} \, N_{2} + {y_{3}\over 2}  \, \phi_3 \,\overline{N_{3}^c} \, N_{3}
+ {\rm h.c.}\right) \nonumber \\
 &+ V_0(\phi_1, \phi_3)\,,
\end{align}
As before, we ignore any mixing between the SM Higgs doublet $\widetilde{\Phi}$ with the complex scalars $\phi_1, \phi_3$. Here $\phi_3$ couples only to the RH neutrino $N_3$, whereas $\phi_1$ couples to both $N_1$ and $N_2$.\footnote{In the next section we will further generalise introducing also a field $\phi_2$ coupling independently to $N_2$.} This can be ensured by giving nonzero $U(1)_{L_{1}}$ charges to $N_1$ and $N_2$ and half of their complementary charge to $\phi_1$, whereas $N_3$ and $\phi_3$ have similar complementary charges under $U(1)_{L_3}$ only. Furthermore, we have chosen a basis where $\phi_1$ and $\phi_3$ only couple to the diagonal elements of the RH neutrino mass matrix.

Analogously to the single majoron model, we write the complex fields as $\phi_1 = \varphi_1 e^{i \theta_1}/\sqrt{2}$ and $\phi_3 = \varphi_3 e^{i \theta_3}/\sqrt{2}$ and assume that the vacuum expectation values are along the real axis, $\vev{\phi_1} = v_1/ \sqrt{2}$ and $\vev{\phi_3} = v_3/ \sqrt{2}$. After spontaneous breaking of both $U(1)$ symmetries, 
$J_1 = v_1\delta\theta_1$ and $J_3 = v_3\delta\theta_3$ are identified as two majorons. We further assume the hierarchy $v_3 \gg v_1$, so that the RH neutrino mass spectrum is hierarchical $M_3 \gg M_1 \simeq M_2 \simeq M$.
 
The $U(1)_{L_1} \times U(1)_{L_3}$ symmetry allows the usual quadratic and quartic terms for both $\phi_1$ and $\phi_2$. It also allows a quartic mixing between the two scalars, so that the tree level potential can now be written as
\begin{align}
	V_0(\phi_1, \phi_3) = -{\mu_1^2}|\phi_1|^2 + \lambda_1 |\phi_1|^4 -{\mu_3^2}|\phi_3|^2 + \lambda_3 |\phi_3|^4  + \zeta |\phi_1|^2 |\phi_3|^2. \label{Vzero2}
\end{align}
At sufficiently high temperatures, 
both symmetries are restored. At temperatures $T \sim v_3$, spontaneous breaking of $U(1)_{L_3}$ generates the massless majoron field $J_3$.  
From this first phase transition we can expect a negligible contribution to the GW spectrum at observable frequencies, as we have seen in the previous section.
After the $\phi_3$ phase transition has completed, and $\phi_3$ has settled down to its VEV $v_3$, the $\phi_1$ phase transition starts.   
Interestingly, as we are going to show, the nonzero mixing of $\phi_1$ with $\phi_3$ implies that the tree-level zero-temperature effective potential 
of $\phi_1$ now gains a cubic term, which can make the phase transition of $\phi_1$ strong enough to produce an observable GW spectrum.

Let us then now focus on the phase transition of $\phi_1$ at a lower scale. 
Writing the potential Eq.~\eqref{Vzero2} in terms of the real fields $\varphi_1$, $\varphi_3$, the minimization conditions yield
\begin{align}
	\mu_1^2 &= \lambda_1 \, v_1^2 + \frac{\zeta}{2} v_3^2, \\
	\mu_3^2 &= \lambda_3 \, v_3^2 + \frac{\zeta}{2} v_1^2.
\end{align}
Because of the mixing term in the potential, the scalar mass matrix has non-vanishing off-diagonal terms. Since $\varphi_3$ has already completed the
phase transition, we can write $\varphi_3 = v_3 + \delta\varphi_3$. On the other hand, we have to use the unshifted field $\varphi_1$,
since we want to describe its phase transition. Following \cite{Kehayias:2009tn}, the mass matrix 
can be diagonalized by rotating the basis vectors, so that $\varphi_1$ and $\varphi_3$ 
can be expressed in terms of the new mass eigenstates $\bar{\varphi}_1$ and $\delta\bar{\varphi}_3$ 
\begin{align}
\varphi_1 &= \bar{\varphi}_1 \cos{\theta} - \delta\bar{\varphi}_3\sin{\theta} \label{varphi1}, \\
\varphi_3 &=   v_3 + \bar{\varphi}_1 \sin{\theta} + \delta\bar{\varphi}_3\cos{\theta}\label{varphi3},
\end{align}
where the rotation angle can be determined, assuming $v_3 \gg v_1$, to be
\begin{align}
	\theta \simeq -\frac{\zeta v_1}{2\lambda_3 v_3}. 
\end{align}
\textcolor{black}{In order to see the impact of the mixing term on the phase transition of $\varphi_1$, 
we expand $\varphi_3$ as given by Eq.~\eqref{varphi3} in the potential in Eq.~ (\ref{Vzero2}). The reason for this is that, since the phase transition of $\varphi_3$ has completed by the time $\varphi_1$ undergoes a phase transition, we can expand $\varphi_3$ around its VEV $v_3$. Notice that from Eq.~(\ref{varphi3}), one can see that 
the phase transition of $\varphi_1$ will induce a small shift of the $\varphi_3$ VEV given by $\bar{\varphi}_1 \sin{\theta}$. 
However,  since $\theta \propto v_1/v_3$ is tiny, this has no effect on the $\varphi_1$ phase transition. 
Although the mass eigenstates above are for the tree-level, zero-temperature, shifted fields, we will treat the aforementioned expansions simply as change of basis.}
In this basis, expanding the quartic mixing term in Eq.~\eqref{Vzero2} in terms of the mass eigenstates yields a cubic term for $\bar{\varphi}_1$,
\begin{align} \label{cubic}
	\frac{\zeta}{4}\varphi_1^2 \varphi_3^2 &\xrightarrow{v_3 \gg v_1} -\frac{\zeta^2}{4}\frac{v_1}{\lambda_3} \bar{\varphi}_1^3 + \ldots
\end{align}
\textcolor{black}{The dots denote the presence of additional terms $\propto (v_1/v_3)^2$ that can be neglected.}
Furthermore, to leading order, since the mixing angle $\theta$ is very small, the mass eigenstate $\bar{\varphi}_1$ almost coincides with $\varphi_1$.\footnote{
There are other subleading terms that are suppressed by the small ratio $v_1/v_3$ and can be neglected. For a full derivation of all terms one can
start from Eq.~(\ref{Vzero2}) and rewrite it in terms of $\varphi_1$ and $\varphi_3$ as 
\be
V_0(\varphi_1, \varphi_3) = -{1\over 2}{\mu_1^2}\varphi_1^2 + {\lambda_1\over 4} \varphi_1^4 -{1\over 2}{\mu_3^2}\varphi_3^2 + {\lambda_3\over 4} \varphi_3^4  
+ {\zeta\over 4} \varphi_1^2 \,\varphi_3^2. \label{Vzerophi}
\ee 
One can then rewrite $\varphi_1$ and $\varphi_3$ in terms of $\bar{\varphi}_1$ and $\delta\bar{\varphi}_3$ using Eqs.~(\ref{varphi1}) and (\ref{varphi3}).
It is easy to see that, neglecting ${\cal O}(\theta)$ and ${\cal O}(\delta\bar{\varphi}_3)$ subleading terms, one obtains 
the thermal effective potential Eq.~(\ref{Veff2}) describing the dynamics of $\varphi_1$. 
}
As anticipated,  the net effect is that a non-vanishing zero temperature cubic term appears in the thermal effective potential of $\varphi_1$ that can now be written as
\begin{align}
	V_{\rm eff}(\varphi_1, T) \approx {1\over 2}\, \widetilde{M}_T^2\,\varphi_1^2 - (A \, T + C) \, \varphi_1^3 + \frac{1}{4}\lambda_T\, \varphi_1^4 \,, \label{Veff2}
\end{align}
where $C={\zeta^2v_1}/({4}{\lambda_3})$. Comparing Eq.~\eqref{Veff2} to Eq.~\eqref{VTeffminimal2}, the expressions for $\widetilde{M}_T, A$ and $\lambda_T$ are obtained from Eqs.~\eqref{MTsq}-\eqref{lambdaT} with the replacement $\lambda \rightarrow \lambda_1, v_0 \rightarrow v_1$.  Since the heaviest RH neutrino mass $M_3 \sim v_3 \gg v_1$, we 
can assume that the $N_3$'s have fully decayed at the onset of the $\phi_1$ phase transition. On the other hand,
we can assume that both lighter RH neutrinos are fully thermalised and, therefore, take $N=2$.\footnote{On the other hand, notice that in the
Coleman-Weinberg potential in Eq.~(\ref{V01}) one still has three RH neutrinos. This mismatch would produce a logarithmic term in the effective
thermal potential, as pointed out in \cite{DiBari:2021dri}. However, as it has been shown there and we verified, 
neglecting this term is a very good approximation in the calculation of the GW spectrum.}

\begin{figure}\label{fig:alphabeta}
\centering
\includegraphics[width=0.7\textwidth]{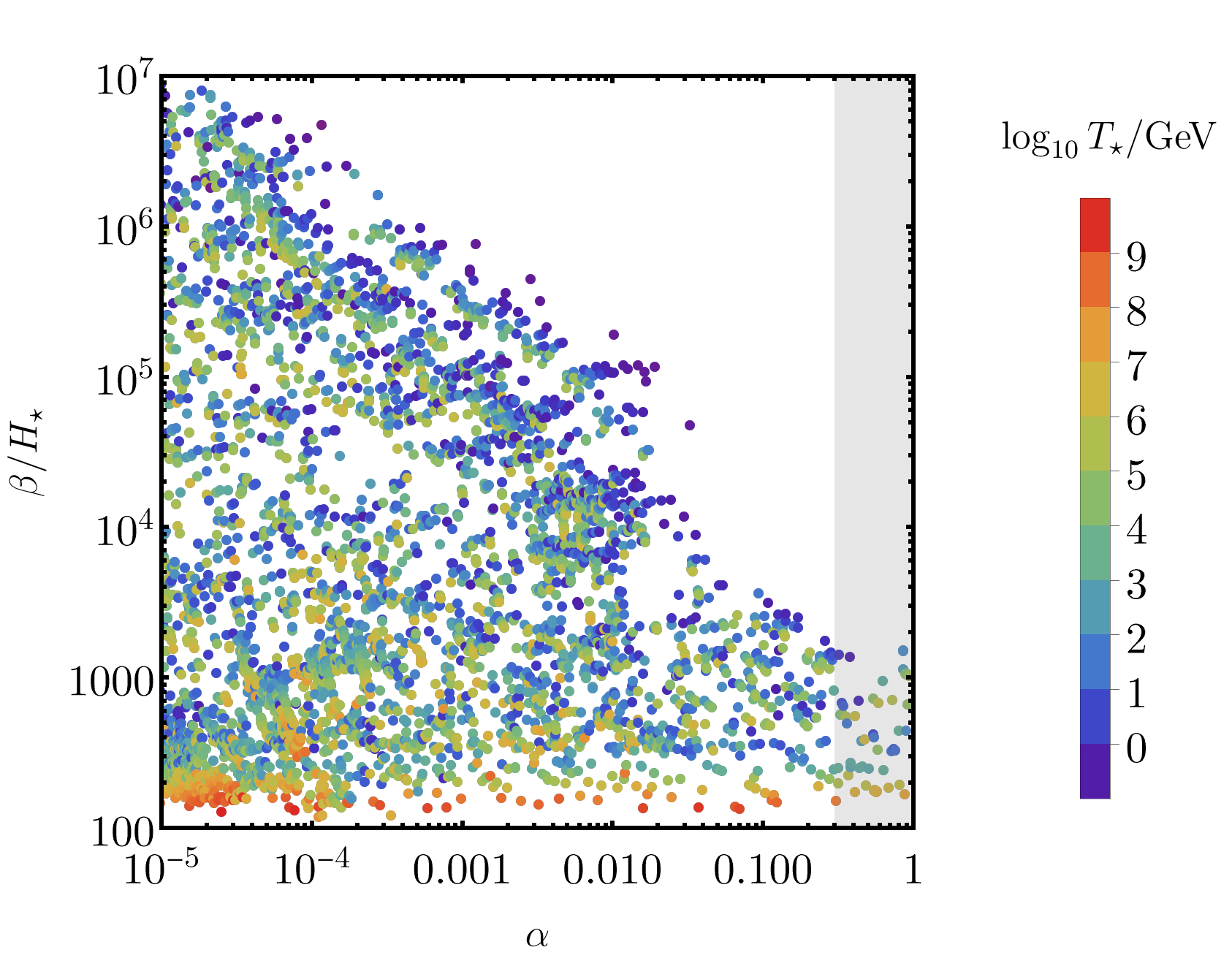}
\caption{GW parameters $\alpha$ and $\beta/H_\star$ from scan over the model parameters $\lambda_1$, $v_1$, $M$ and $C$. The color bar represents $\log_{10}T_\star/{\rm GeV}$ for each point. }
\end{figure}

The cubic term at zero temperature helps to strengthen the phase transition of $\phi_1$. To illustrate this, we perform a random scan over the model parameters $(\lambda_1, v_1, C)$\footnote{\textcolor{black}{Strictly speaking, the model parameters are the coefficients that appear in the potential Eq.~\eqref{Vzero2}, namely, $\mu_1$, $\lambda_1$, $\mu_2$, $\lambda_2$ and $\zeta$. $v_1$, $v_3$ and $C$ can be expressed in terms of these parameters. We choose $\lambda_1$, $v_1$, $\lambda_3$, $v_3$ and $C$ as free parameters.}} in the range $10^{-6} \leq \lambda_1 \leq 1$, $1 \leq v_1/\text{GeV} \leq 10^7$, $10^{-4} \leq M/v_1 \leq 10$ and $10^{-8} \leq C/v_1 \leq 1$ and calculate the GW parameters $T_\star$, $\alpha$ and $\beta/H_\star$, following section~\ref{GWFOPT}. In Fig.~\ref{fig:alphabeta} we show the results of the scan, where the color map represents $\log_{10}T_\star/{\rm GeV}$ at each point. The model allows $\alpha \gtrsim \mc O(1)$ and $\beta \gtrsim 10^7$, however, as we discussed, 
we consider only points for $\alpha \leq 0.3$ and $\beta/H_{\star} > 100$.

\begin{figure}\label{fig:TpAlphaBeta}
\centering
\includegraphics[width=\textwidth]{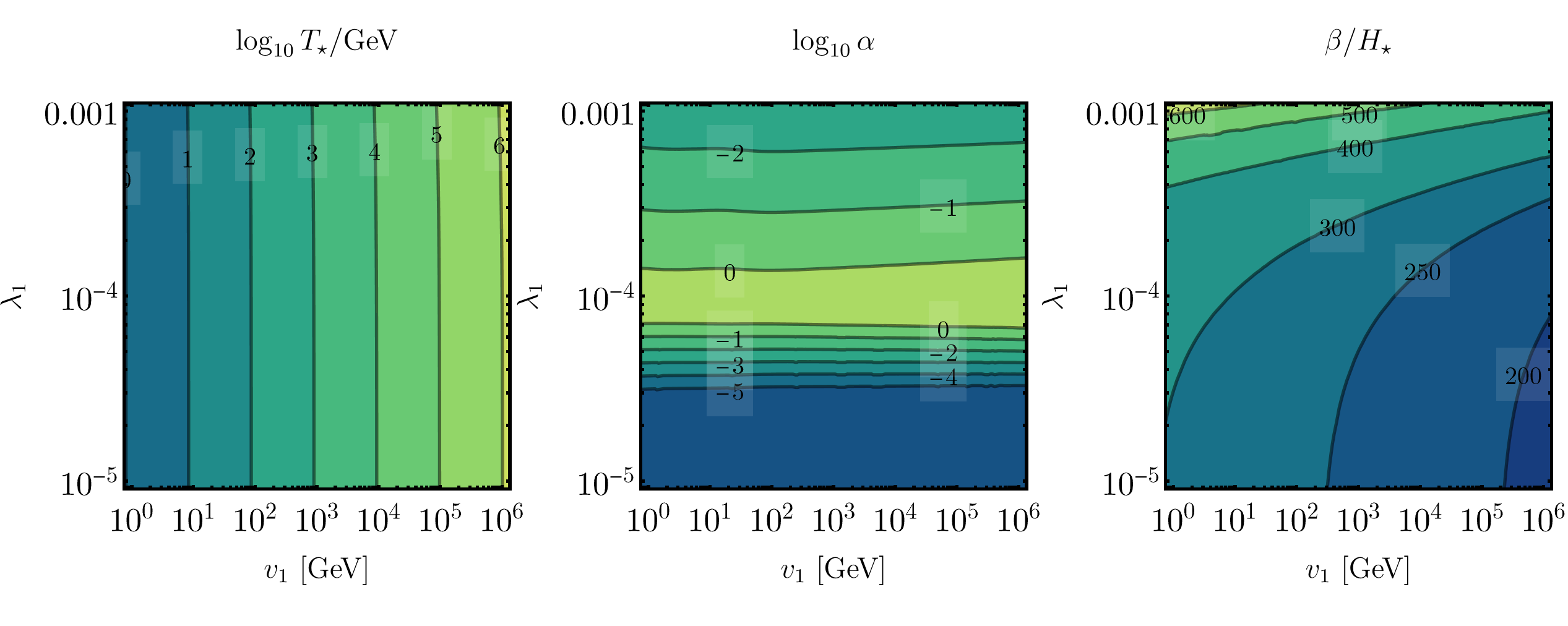}
\caption{Dependence of the GW parameters $T_\star$, $\alpha$ and $\beta/H_\star$ on the model parameters $\lambda_1$ and $v_1$, setting $M= 0.15 \, v_1$ and $C = 0.002 \, v_1$. }
\end{figure}

To get a better understanding of how $T_\star$, $\alpha$ and $\beta/H_\star$ depend on the model parameters, we look at a two-dimensional slice of the parameter space in terms of $\{\lambda_1, v_1\}$ setting $M= 0.15 v_1$ and $C = 0.002 v_1$. The results are shown in Fig.~\ref{fig:TpAlphaBeta}. We find that $T_\star \sim \mc O(v_1)$ and is nearly independent of $\lambda_1$. On the other hand $\alpha$ is essentially determined by $\lambda_1$, and peaks near $\lambda_1 \sim \mc O(10^{-4})$. Finally, $\beta/H_\star$ depends on both $\lambda_1$ and $v_1$.

We now look at the gravitational wave spectrum for the three benchmark points listed in Table~\ref{table:BP}.\footnote{\textcolor{black}{The corresponding values of the parameters $\mu_1$, $\mu_3$ and $\zeta$ are as follows, assuming $v_3 = 10^{14}$ GeV and $\lambda_3 = 0.001$: \Circled{A}: $\mu_1 = 6.5 \times 10^{12} $ GeV, $\mu_3 = 3.2\times 10^{12}$ GeV, $\zeta = 0.0084$; \Circled{B}: $\mu_1 = 6.0 \times 10^{12} $ GeV, $\mu_3 = 1.0\times 10^{12}$ GeV, $\zeta = 0.0072$; \Circled{C}: $\mu_1 = 1.2 \times 10^{13} $ GeV, $\mu_3 = 1.0\times 10^{12}$ GeV, $\zeta = 0.0285$.}} The benchmark points have been chosen to maximize the GW amplitude from first order phase transition in their respective peak frequencies. The resulting signals are shown in Fig.~4, along with the GW spectrum from the global cosmic strings for $v_3 = 10^{15}$, $5 \times 10^{14}$, $2 \times 10^{14}$ and $10^{14}$ GeV. \textcolor{black}{As expected, the dominant contribution is from sound waves (thin dashed lines), whereas the MHD turbulence (thin dot-dashed lines) contributes to the high-frequency tail of the spectra.}  The peak amplitude of these signals are consistent with what one would expect from the range of $\alpha$ and $\beta/H_\star$ where our calculation of GW spectrum is valid, as discussed in Appendix \ref{app:A}.

\begin{table}
\centering
{\renewcommand{\arraystretch}{1.2}
\begin{tabular}{c@{\hskip 0.1in}c@{\hskip 0.1in}c@{\hskip 0.1in}c@{\hskip 0.1in}c@{\hskip 0.1in}|c@{\hskip 0.1in}c@{\hskip 0.1in}c@{\hskip 0.1in}c}
    \toprule

  B.P.  & $\lambda_1$ & $v_1$ [GeV] & $M$ [GeV] & $C$ [GeV] & $\alpha$ & $\beta/H_\star$ & $T_{\star}$ [GeV] & $\langle \varphi_1\rangle^{\rm true}_{T_{\star}}$ [GeV] \\
 \midrule
\Circled{A} & $0.00057$ & $1188.22$  & $186.53$ & $20.79$ & $0.29$ & $244.65$ & $5863.12$ & $1.38 \times 10^5$ \\
\Circled{B} & $0.00061$ & $2.32 \times 10^5$  & $3.63 \times 10^4$ & $3023.02$ & $0.30$ & $204.66$  & $7.81 \times 10^5$ & $1.79 \times 10^7$\\
\Circled{C} & $0.00036$ & $9.88 \times 10^6$  & $1.08 \times 10^6$ & $2 \times 10^6$ & $0.30$ & $141.48$ & $7.51 \times 10^{8}$ & $1.92 \times 10^{10}$ \\
\bottomrule
\end{tabular}
}
\caption{Benchmark points for gravitational wave signals from first order phase transition of $\varphi_1$.}
\label{table:BP}
\end{table}


\begin{figure} 
    \centering
      \includegraphics[width=0.99\textwidth]{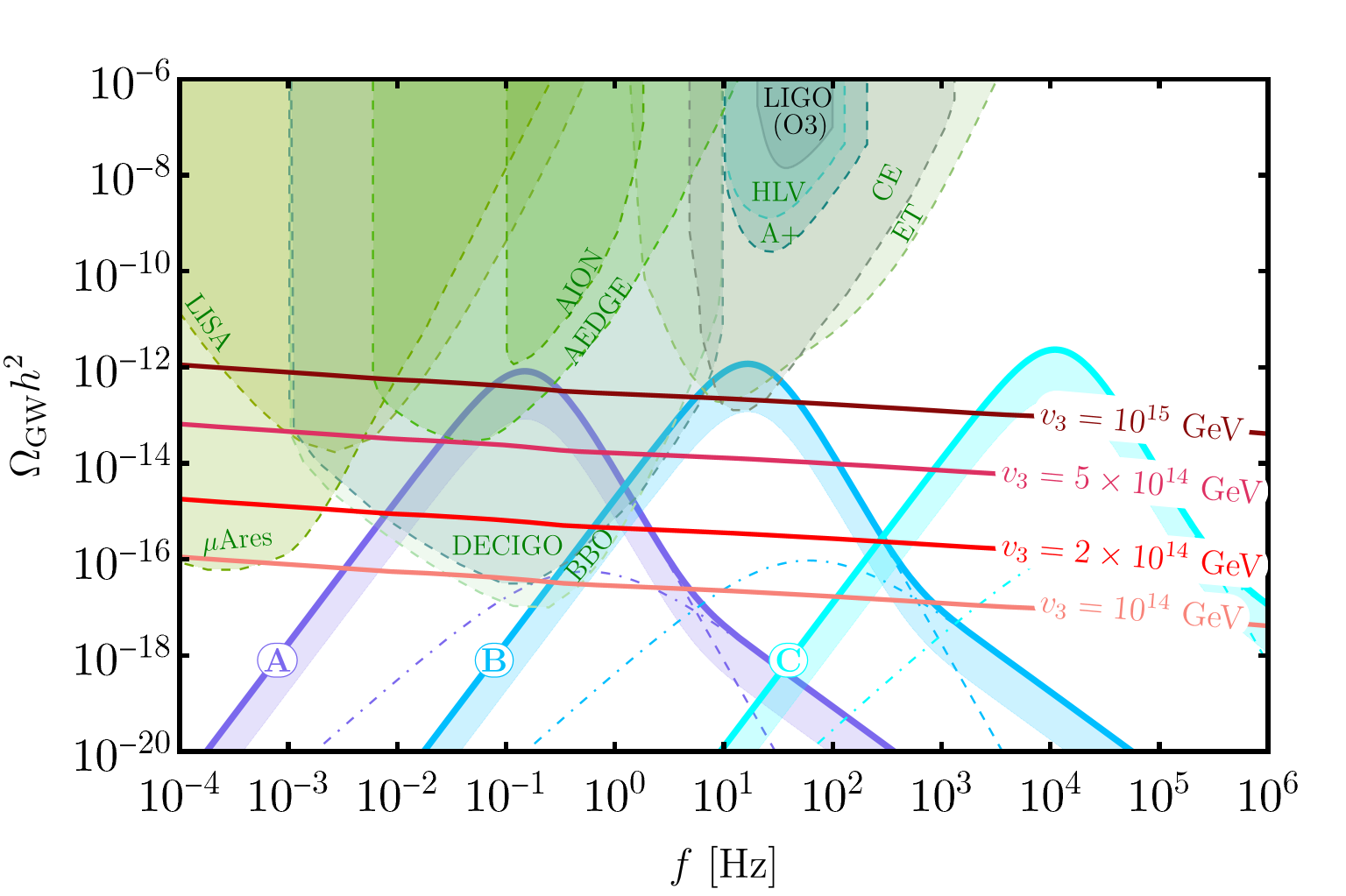}
    \label{fig:AllBenchmarkPoints}
    \caption{Gravitational wave spectrum from first order phase transition of $v_1$ for three benchmark points shown in Table~\ref{table:BP} and from global cosmic string formed by spontaneous breaking of the global $U(1)_{L_3}$ symmetry by $\phi_3$ for four representative cases. The shaded region in the phase transition signals represent uncertainties in the calculation of the gravitational wave amplitude. Sensitivities and upper bounds from various upcoming and present interferometers are also shown.
See main text for more details.}
\end{figure}

The peak amplitude of the benchmark points \Circled{A} and \Circled{B} are sensitive to  DECIGO, BBO, AEDGE, and ET, CE, respectively, while point \Circled{C} peaks at a higher frequency. In all cases, the peak amplitude is larger than the global cosmic string induced spectrum for $v_3 \lesssim 10^{15}$ GeV. For any benchmark point and a given $v_3$, the combined gravitational wave spectrum would look like a peak towering above the slightly tilted plateau.\footnote{However, if $v_1 < T_{\rm RH} < v_3$, we would only have the phase transition signal. The signal is still enhanced since $\phi_3$ gets  a VEV prior to the phase transition of $\phi_1$, although no symmetry breaking appears near the scale $v_3$. For $T_{\rm RH} < v_1$, even the phase transition signal would disappear as the universe is in a broken phase at $T_c \sim v_1$.} While the wideband nature of the global cosmic string induced signal offers detection possibility at multiple interferometers, the larger peak from first order phase transition provides better visibility. Combining the two features, a unique gravitational wave signal emerges for models with two scalars, one breaking a global $U(1)$ symmetry at ultraviolet scales and the other undergoing a strong first order phase transition at lower scales.\footnote{GW spectra where both contributions from cosmic strings and phase transition combined together were also found in \cite{Fornal:2020esl,Bosch:2023spa}.}

\section{GW from Majorana mass genesis in a three-majoron model}
A straightforward generalization of the model is to include three complex scalars with hierarchical VEVs, so that each scalar gives mass to one of the RH neutrinos,
\begin{align}
-\mc L_{N_I+\phi_I} &\supset  \left(\overline{L_a}h_{a I} H N_I + \frac{y_{1}}{2}\phi_1 \overline{N^c_1} N_1 + \frac{y_{2}}{2}\phi_2 \overline{N^c_2} N_2 + \frac{y_{3}}{2}\phi_3 \overline{N^c_3} N_3 +\text{h.c.} \right)
 + V_0(\phi_1, \phi_2, \phi_3). 
\end{align}
The Lagrangian has a $U(1)_{L_1} \times U(1)_{L_2} \times U(1)_{L_3}$ symmetry, with each $U(1)$ corresponding to each scalar.\footnote{\textcolor{black}{Here we do \emph{not} identify $\overline{L_a}$ with the flavor eigenstates ($a = e, \mu, \tau$). Then, realistic lepton mixing arises when the Yukawa matrix $h_{aI}$ is rotated to the flavor basis.}}
We denote the VEVs as $\vev{\phi_I} \equiv v_I$ and without loss of generality assume $v_3 \gg v_2 \gg v_1$. The tree-level scalar potential is given by
\begin{align}\label{V3scalar}
V_0 (\phi_1, \phi_2, \phi_3) &= \sum_{I=1,2,3} \left[-\mu_I^2 \phi_I^* \phi_I + \lambda_I (\phi_I^* \phi_I)^2\right] + \sum_{I,J,I\neq J}^{1,2,3} \frac{\zeta_{IJ}}{2} (\phi_I^* \phi_I)(\phi_J^* \phi_J).
\end{align}
After spontaneous breaking of the global $U(1)$ symmetries, the three RH neutrinos get nonzero Majorana mass from the VEV of $\phi_1$, $\phi_2$, $\phi_3$. Assuming these VEVs are along the radial axis, one can identify the phase part of the complex scalars as massless majorons.

The mixing terms $\zeta_{IJ}$ in Eq.~\eqref{V3scalar} introduce a zero temperature cubic term to the effective potential of a scalar with smaller VEV. 
As before, the phase transition of $\phi_3$ occurring at around the scale $v_3$ is not expected to generate any strong gravitational wave signal, since there is no zero temperature cubic term in its effective potential. However, the spontaneous breaking of $U(1)_{L_3}$ at this scale would generate global cosmic string induced gravitational waves, which can be probed if $v_3 \gtrsim 10^{14}$ GeV. Suppose the phase transition of $\phi_3$ is completed before the universe cools down to the scale $v_2$, when $\phi_2$ undergoes a phase transition. The quartic mixing of $\phi_2$ with $\phi_3$ now introduces a zero temperature cubic term to the thermal effective potential of $\phi_2$, resulting in a strong first order phase transition and associated gravitational wave from the sound waves. At this stage $\phi_1$ does not play any role in the phase transition of $\phi_2$. Then, during the phase transition of $\phi_1$ at around the scale $v_1$, the other two scalars have already completed their phase transition and together they would introduce an effective zero temperature cubic term from their mixing with $\phi_1$, resulting in a strong phase transition and subsequent gravitational wave signal.

\textcolor{black}{Because of the assumed hierarchy $v_3 \gg v_2 \gg v_1$, the cubic terms for the phase transition of $\phi_2$ and $\phi_1$ depend predominantly on the mixing parameters $\zeta_{23}$ and $\zeta_{12}$, respectively, $C_2 \simeq v_2 \zeta_{23}/(4\lambda_3)$ and $C_1 \simeq v_1 \zeta_{12}/(4\lambda_2)$. The mass parameters $\mu_I$ can be expressed from minimization of the zero-temperature tree-level potential as
\begin{align}
    \mu_I^2 = \lambda_I v_I^2 + \sum_{J\neq I} \frac{\zeta_{IJ}}{2}v_J^2.
\end{align}
As before, we will treat the symmetry breaking scales $v_I$ and the cubic terms $C_I$ as model parameters instead of $\mu_I$ and $\zeta_{IJ}$, since the latter can be determined from the former in conjugation with the rest of the parameters $\lambda_I$.
}

\begin{table}
\centering
{\renewcommand{\arraystretch}{1.2}
\begin{tabular}{c@{\hskip 0.1in}c@{\hskip 0.1in}c@{\hskip 0.1in}c@{\hskip 0.1in}c@{\hskip 0.1in}|c@{\hskip 0.1in}c@{\hskip 0.1in}c@{\hskip 0.1in}c}
    \toprule

    & $\lambda_I$ & $v_I$ [GeV] & $M_I$ [GeV] & $C_I$ [GeV] & $\alpha$ & $\beta/H_\star$ & $T_{\star}$ [GeV] & $\langle \varphi_I\rangle^{\rm true}_{T_{\star}}$ [GeV] \\
 \midrule
 \Circled{D} & $0.00027$ & $1188.2$  & $186.5$ & $10.79$ & $0.30$ & $241.37$ & $5196.52$ & $1.50 \times 10^5$ \\
 \Circled{E} & $0.00029$ & $2.32 \times 10^5$  & $3.63 \times 10^4$ & $1523.02$ & $0.30$ & $203.53$ & $6.7 \times 10^5$ & $1.88 \times 10^7$ \\
\bottomrule
\end{tabular}
}
\caption{Benchmark point for gravitational wave signal from first order phase transition of $\varphi_1$, denoted by \Circled{D} ($I=1$) and $\varphi_2$, denoted by \Circled{E} ($I=2$). \textcolor{black}{Taken together, the parameters given in this table constitute \emph{one} benchmark point for the three-majoron model.}}
\label{table:BP2}
\end{table}

\begin{figure} 
    \centering
      \includegraphics[width=0.89\textwidth]{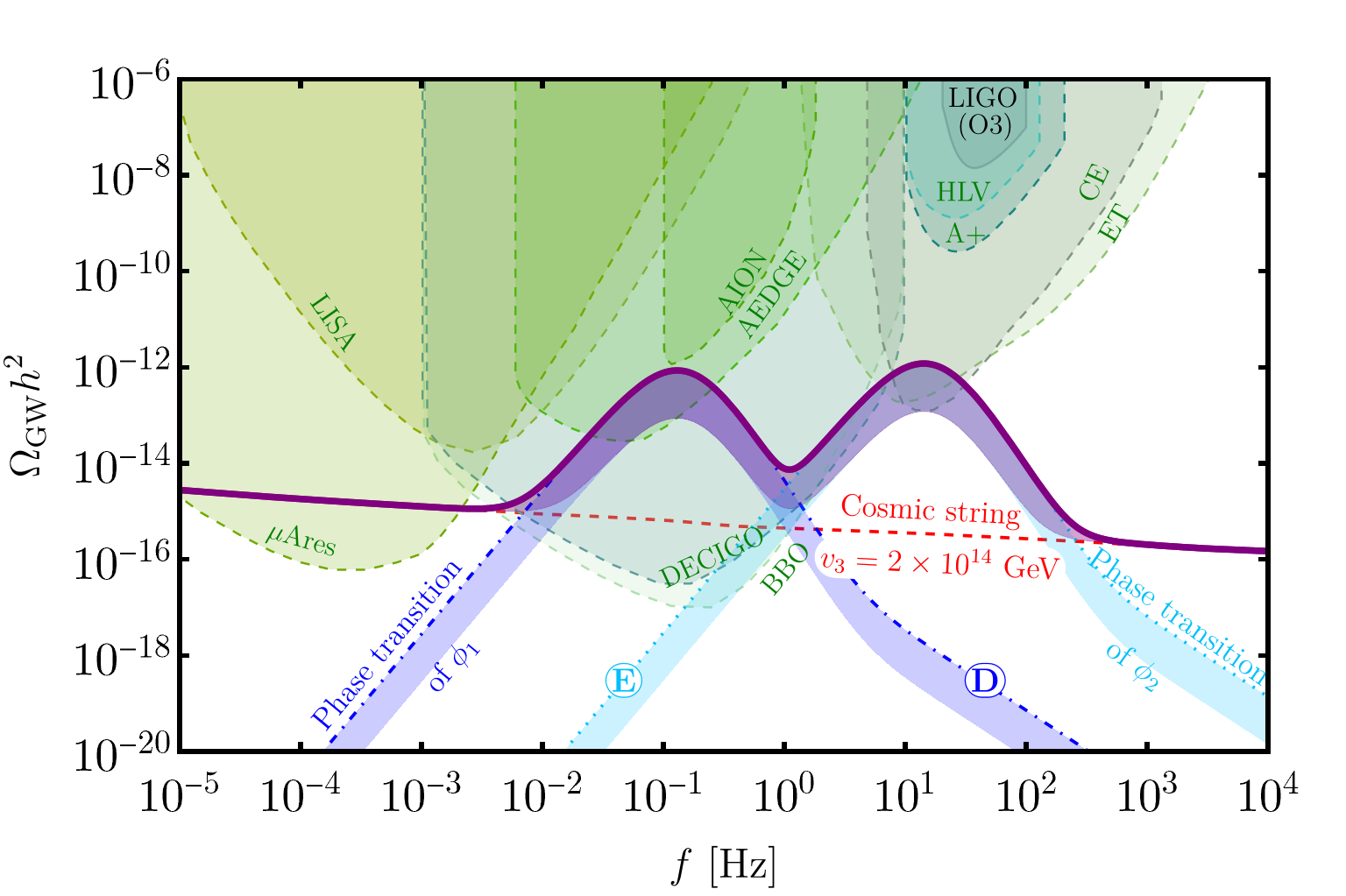}
    \label{fig:threeGW}
    \caption{Gravitational wave spectrum from first order phase transition of $\phi_1$ and $\phi_2$, with corresponding parameters shown in rows \Circled{D} and \Circled{E}, respectively, in Table~\ref{table:BP2}, and from  cosmic string formed by spontaneous breaking of the global $U(1)_{L_3}$ symmetry by $\phi_3$ at $v_3 = 2\times 10^{14}$ GeV. Individual GW contributions are shown with dotted, dashed and dotdashed lines, while the combined spectrum is shown with a solid curve. Band in GW spectrum from phase transition represent the possible $\mc O(0.1)$ suppression in the parameter $\widetilde{\Omega}_{\rm gw}$.}
\end{figure}

Typically the percolation temperature $T_\star$ is proportional to the VEV of the corresponding scalar undergoing the phase transition. From Eq.~\eqref{fpeak}, this implies that the combined effect of the phase transition of the three scalars may yield a double peaked gravitational wave spectrum, with one peak at a lower frequency due to the phase transition of $\phi_1$, and another peak at a higher frequency due to the phase transition of 
$\phi_2$. Together with a global cosmic string induced gravitational wave spectrum from $U(1)_{L_3}$ breaking, the combined amplitude of the gravitational wave signal may resemble twin peaks over a slightly slanted plateau, if the phase transition signals are sufficiently strong.\footnote{This is, of course, assuming $T_{\rm RH} > v_3$, otherwise the signals that can be generated only above a given $T_{RH}$ would not appear.} In Table~\ref{table:BP2}, we show a benchmark point consisting of the phase transition of $\phi_1$, denoted by \Circled{D} and the phase transition of $\phi_2$, denoted by \Circled{E}, that together with $v_3 = 2\times 10^{14}$ GeV generate the combined gravitational wave signal shown in Fig.~\ref{fig:threeGW}\footnote{\textcolor{black}{The corresponding model parameters are $\zeta_{12} = 0.0032$, $\zeta_{23} = 0.0051$, $\mu_1 = 4.47 \times 10^{12}$ GeV, $\mu_2 = 1.01 \times 10^{13}$ GeV, $\mu_3 = 6.32 \times 10^{12}$ GeV, taking $\lambda_3 = 0.001$ and $\zeta_{13} = 0.001$.}}. Notice that in this case we have assumed that at each phase transition $N=1$, corresponding to a situation where only the RH neutrino species $N_I$,
coupling to its associated scalar field $\phi_I$ undergoing the phase transition, 
is fully thermalised, while the other two either have fully decayed 
or have not yet thermalised. This assumption is quite natural because of the strong hierarchy we are assuming for the
$v_I$'s, implying that a strong hierarchy of the RH neutrino mass spectrum and in turn of the equilibration
temperatures (see Eq.~(\ref{Teq}). 
\section{Conclusion}
We have investigated the gravitational wave signatures of the majoron model of neutrino mass generation and have identified two sources of gravitational waves. In the simplest single majoron model, a complex scalar couples to the RH neutrinos and generates their Majorana mass after spontaneously breaking the global lepton number symmetry. The breaking of a global symmetry creates global cosmic strings which can produce gravitational waves, with a different spectrum as compared to that from local Nambu-Goto strings. In the observable frequency window, the amplitude of this signal mildly declines as $\log^3[{1/f}]$, and still remains sensitive to upcoming GW interferometers if the symmetry is broken at a scale in between $10^{14}-10^{15}$ GeV. However, there is a possible additional source of GWs in this model, since the complex scalar gets a nonzero vacuum expectation value and in the process might undergo a first order phase transition. If such a phase transition is sufficiently strong, it could generate a peaked GW signal which may tower over the global cosmic string signal.

For the simplest model with just one complex scalar coupling to all three RH neutrinos, we confirm the result of Ref.~\cite{DiBari:2021dri} that the phase transition signal is too feeble to be detected. However, we point out that the global cosmic string signal even in this model can be detected if the lepton number symmetry is broken at around $10^{14}-10^{15}$ GeV.

We then considered an extended majoron model, introducing two complex scalars with hierarchical vacuum expectation values, one giving mass to the heaviest RH neutrino and the other to the remaining two lighter ones. Assuming the scalars are charged under separate lepton number symmetries and have a quartic mixing between them, we explored the global cosmic string induced GW spectrum which is generated when the heaviest RH neutrino gets a mass. We showed that, while the phase transition of the associated scalar remains weak, its mixing with the other scalar introduces a zero-temperature cubic term to the potential of the latter, and greatly enhances the GW signal from its phase transition. We have discussed examples where the combined GW spectrum of the model may have an observable bump or peak due to the phase transition signal, visible in the slanted plateau region from the cosmic string signal, where such a bump may appear anywhere over the whole range of observable frequencies.

Finally, we have discussed an interesting possibility of a double peaked spectrum which may occur over the global cosmic string plateau region, where such a spectrum may arise from an extension of the majoron model to include three complex scalars. This rather plausible model is easily implemented for a hierarchical RH neutrino mass spectrum, where each RH Neutrino gets its mass from the spontaneous breaking of its respective lepton number symmetry. Such a double peaked spectrum provides a characteristic signature of the three majoron model of neutrino mass generation. 

We have also noticed how the observation of such a GW spectrum would give us a precious information on the cosmological history and in particular on the reheating temperature of the universe. We have implicitly assumed that  this was higher than all vacuum expectation values and critical temperatures so that the GW spectra are produced through 
the entire range of corresponding frequencies. However, if the reheating temperature is below the vacuum expectation value of one of the complex scalar fields, then the phase transition would not take place and the signal would be absent. At the same time it should be mentioned that the model we have presented can be clearly combined with (minimal) leptogenesis \cite{Fukugita:1986hr} since the decays of the RH neutrinos would produce a $B-L$ asymmetry that can then be partly converted
into a baryon asymmetry. Therefore, the observation of the GW spectra in this model would also provide a strong test of leptogenesis. Moreover, as proposed in \cite{DiBari:2020bvn}, a phase transition of the complex scalar field can be also associated to the production of a dark RH neutrino playing the role of dark matter \cite{Anisimov:2008gg}. Future GW experiments have then the  potential to shed light on neutrino mass genesis, cosmological history and origin of matter of the universe.  

\section*{Acknowledgments} 

We acknowledge financial support from the STFC Consolidated Grant ST/T000775/1 
and from the European Union's Horizon 2020 Research and Innovation
Programme under Marie Sk\l odowska-Curie grant agreement HIDDeN European
ITN project (H2020-MSCA-ITN-2019//860881-HIDDeN). We acknowledge the use of the IRIDIS High-Performance Computing Facility and associated support services at the University of Southampton in the completion of this work.
SFK would like to thank IFIC, Valencia for its hospitality. 
We wish to thank Chia-Feng Chang, Yanou Cui, Nikolai Husung, Shaikh Saad, Graham White and David Weir
for useful discussions. 

\appendix

\section{Dependence of peak GW amplitude on FOPT parameters}\label{app:A}
\begin{figure}\label{fig:alphabetaGWpeak}
\centering
\includegraphics[width=0.7\textwidth]{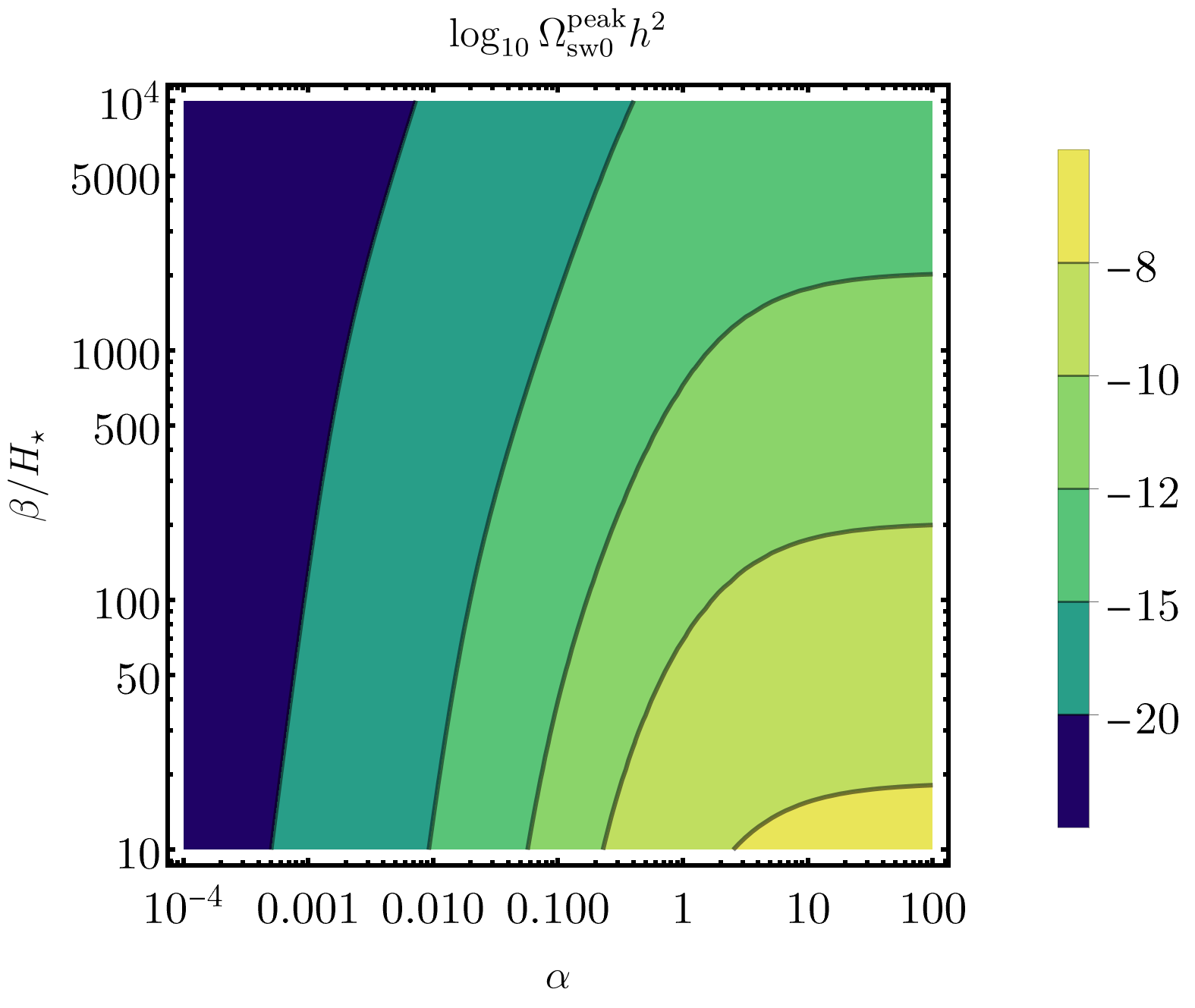}
\caption{Contours of the peak GW amplitude from sound waves as a function of $\alpha$ and $\beta/H_\star$.}
\end{figure}

The peak of the GW amplitude from sound waves can be expressed as  a function of FOPT parameters $\alpha$ and $\beta/H_\star$, as seen from Eq.~\eqref{omegasw}. In Fig.~\ref{fig:alphabetaGWpeak} we show contours of $\log_{10}{\Omega_{\rm sw0}^{\rm peak}h^2}$. This plot shows that typically the peak amplitude of GW sourced by sound waves would be weaker than $10^{-11}$, as we have seen in the benchmark points of Figs.~\ref{fig:AllBenchmarkPoints} and \ref{fig:threeGW}.
\newpage

\bibliography{references}
\newpage
\bibliographystyle{JHEP}

\end{document}